\documentstyle[psfig,epsfig]{mn}

\newcommand{\HI}{H\,{\sc i}}
\newcommand{\HII}{H\,{\sc ii}}
\newcommand{\ms}{~m\,s$^{-1}$}
\newcommand{\kms}{~km\,s$^{-1}$}

\newcommand{\Msun}{~M$_{\odot}$}
\newcommand{\tsys}{$T_{\rm sys}$}

\title[The Australia Telescope Compact Array Broadband Backend (CABB)] 
      {The Australia Telescope Compact Array Broadband Backend (CABB):
       Description \& First Results\thanks{The Australia Telescope 
           Compact Array is part of the Australia Telescope National Facility 
           which is funded by the Commonwealth of Australia for operation as 
           a National Facility managed by CSIRO.}  }

\author[Wilson et al.]
       {Warwick E. Wilson$^1$, R.H. Ferris$^1$, P. Axtens$^1$, A. Brown$^1$, 
        E. Davis$^1$, G. Hampson$^1$, \and M. Leach$^1$, P. Roberts$^1$, S. 
        Saunders$^1$, B.S. Koribalski$^{1,8}$, J.L. Caswell$^1$, E. Lenc$^1$, 
        J. \and Stevens$^1$, M.A. Voronkov$^1$, M.H. Wieringa$^1$, 
        K. Brooks$^1$, P.G. Edwards$^1$, R.D. Ekers$^1$, \and B. Emonts$^1$, 
        L. Hindson$^{2,1}$, S. Johnston$^1$, S.T. Maddison$^3$, E.K. 
        Mahony$^{4,1}$, S.S. \and Malu$^7$, M. Massardi$^5$, M.Y. Mao$^{6,1}$, 
        D. McConnell$^1$, R.P. Norris$^1$, D. Schnitzeler$^1$, \and R.
        Subrahmanyan$^7$, J.S. Urquhart$^1$, M.A. Thompson$^2$, and R.M. 
        Wark$^1$  \\
        $^1$CSIRO Astronomy \& Space Science, Australia Telescope National 
            Facility, P.O. Box 76, Epping, NSW 1710, Australia \\
        $^2$Centre of Astrophysics Research, University of Hertfordshire, 
            College Lane, Hatfield AL10 9AB, U.K. \\
        $^3$Centre of Astrophysics \& Supercomputing, Swinburne University of
            Technology, Hawthorn, VIC 3122, Australia \\ 
        $^4$Sydney Institute for Astronomy, School of Physics, University 
            of Sydney, NSW 2006, Australia \\
        $^5$INAF, Osservatorio Astronomico di Padova, Vicolo dell'Osservatorio
            5, I-35122 Padova, Italy \\
        $^6$School of Mathematics and Physics, University of Tasmania, Private
            Bag 37, Hobart 7001, Australia \\
        $^7$Raman Research Institute, Sadashivanagar, Bangalore 560 080, 
            India \\
        $^8$Contact email: Baerbel.Koribalski@csiro.au 
}

\date{Received date; accepted date}

\begin{document}

\maketitle

\begin{abstract}
Here we describe the Compact Array Broadband Backend (CABB) and present first 
results obtained with the upgraded Australia Telescope Compact Array (ATCA). 
The 16-fold increase in observing bandwidth, from $2 \times 128$~MHz to $2 
\times 2048$~MHz, high bit sampling, and addition of 16 {\em zoom} windows 
(each divided into a further 2048 channels) provide major improvements for
all ATCA observations. The benefits of the new system are: (1) hugely 
increased radio continuum and polarization sensitivity as well as image 
fidelity, (2) substantially improved capability to search for and map emission 
and absorption lines over large velocity ranges, (3) simultaneous multi-line 
and continuum observations, (4) increased sensitivity, survey speed and dynamic 
range due to high-bit sampling, and (5) high velocity resolution, while 
maintaining full polarization output. The new CABB system encourages all
observers to make use of both spectral line and continuum data to achieve 
their full potential. 

Given the dramatic increase of the ATCA capabilities in all bands (ranging 
from 1.1 to 105~GHz) CABB enables scientific projects that were not feasible 
before the upgrade, such as simultaneous observations of multiple spectral 
lines, on-the-fly mapping, fast follow-up of radio transients (e.g., the 
radio afterglow of new supernovae) and maser observations at high velocity 
resolution and full polarization.
The first science results presented here include wide-band spectra, high 
dynamic-range images, and polarization measurements, highlighting the 
increased capability and discovery potential of the ATCA. 
\end{abstract}


\begin{figure*} 
 \mbox{\psfig{file=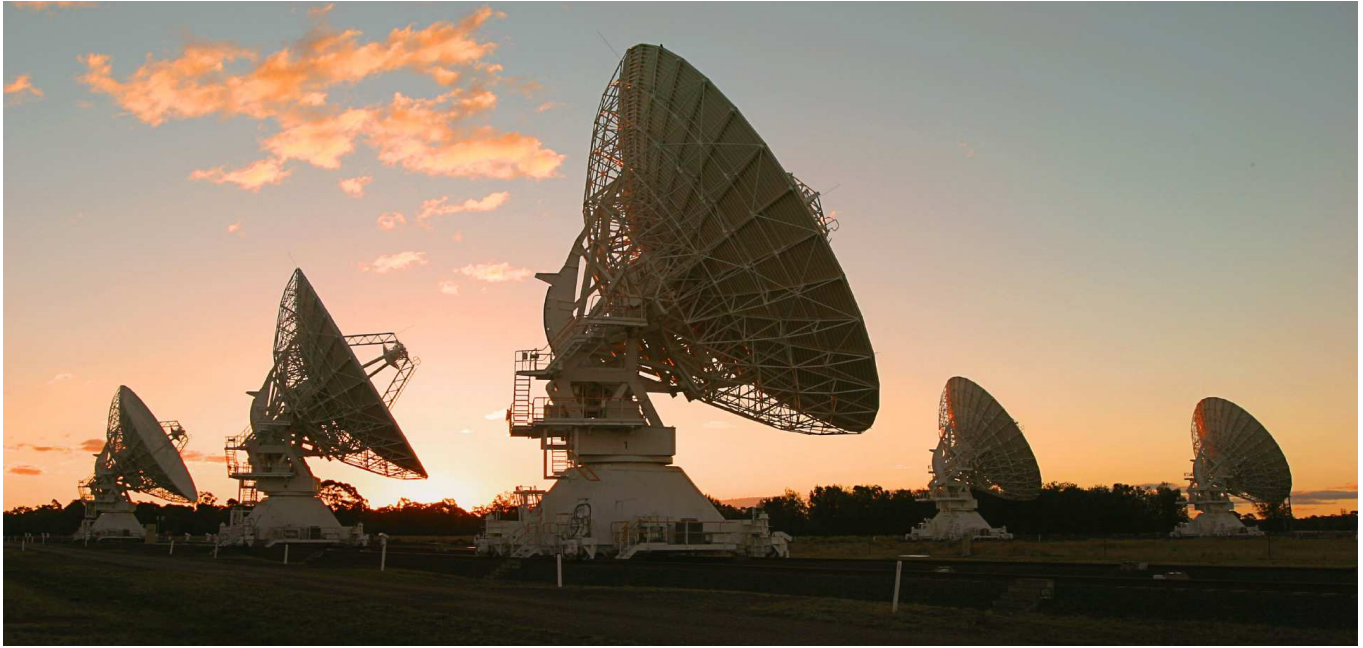,width=18cm,angle=0}} 
\caption{\label{atca1}
  The Australia Telescope Compact Array (ATCA), located at the Paul Wild 
  Observatory near Narrabri, some 550~km northwest of Sydney. Displayed 
  are five of the six 22-m Cassegrain antennas, here arranged in one of 
  the hybrid configurations which include two antennas on the North-South 
  spur. The latter was opened on 26 Nov 1998, just over ten years after the 
  ATCA opening on 2 Sep 1988.}
\end{figure*}

\section{Introduction} 

The Australia Telescope National Facility (ATNF) provides open access to a 
large range of radio telescopes: the 64-m Parkes telescope, the Australia 
Telescope Compact Array (ATCA), the 22-m Mopra telescope, and the Long 
Baseline Array (LBA); see Frater, Brooks \& Whiteoak (1992). Furthermore, 
CSIRO's `Australian SKA Pathfinder' (ASKAP), consisting of $36 \times 12$-m 
dishes, is currently under construction in Boolardy, Western Australia. With 
its expected 30 square degrees field of view at 1.4~GHz, provided by novel 
phased array feeds, ASKAP will be a fast 21-cm survey machine, designed to 
carry out large-scale \HI\ spectral line, transients and radio continuum 
surveys of the sky (Johnston et al.  2007, 2008). 

In this paper we focus on the ATCA and its recent upgrade in bandwidth from 
$2 \times 128$~MHz to $2 \times 2048$~MHz. While greatly increasing the 
sensitivity of continuum and spectral line observations (a factor 16 in 
bandwidth gives a factor four in continuum sensitivity) or, alternately, 
reducing the required observing time for a given continuum sensitivity by 
a factor 16, this upgrade also enables simultaneous (full Stokes) multi-line 
observations, on-the-fly mapping, as well as searching for emission and 
absorption lines over large velocity ranges. The high-bit sampling further 
improves sensitivity, mapping speed and dynamic range for radio frequency 
interference (RFI) suppression.

\subsection{The Australia Telescope Compact Array} 

The ATCA is a radio interferometer consisting of six 22-m dishes (Frater \&
Brooks 1992), creating 15 baselines in a single configuration. While five 
antennas (CA01 to CA05) are movable along a 3-km long east-west track and 
a 214-m long north-south spur (see Figs.~1 \& 2), allowing the creation of 
hybrid arrays\footnote{The ATCA hybrid configurations (H75, H168, H214) 
    provide excellent $uv$-coverage over a 6-h integration time, thereby 
    avoiding the need to observe astronomical sources at low elevations 
    (highly beneficial for mm-wave observations where the earth atmosphere 
    adds substantially to the system temperature and consequently the r.m.s. 
    noise).},
one antenna (CA06) is fixed at a distance of 3-km from the western end of the 
track. Each antenna currently has a set of six cryogenically cooled low noise 
receivers sampling the frequency range from 1.1 to 105~GHz (i.e., wavelengths 
from 3-mm to 30-cm), apart from CA06 which does not have a 3-mm receiver 
system (see Table~1). --- A comprehensive description of aperture synthesis 
and radio interferometry techniques is given in the expert lectures at the 
regularly held Radio Synthesis Schools (e.g., Perley, Schwab \& Bridle 1988, 
and Taylor, Carilli \& Perley 1998) and also Thompson, Moran, \& Swenson 
(2002). \\

In their overview of the Australia Telescope, Frater, Brooks \& Whiteoak (1992)
stated the aim to operate the array in {\em ten} frequency bands, from 0.3 to 
116~GHz. Initially, four bands (1.5, 2.3, 5.0 and 8.6~GHz) were available 
(James 1992; Sinclair et al. 1992) delivering excellent science (Ekers \& 
Whiteoak 1992). The first stage of the millimeter-wave upgrade (Hall et al. 
1997; Koribalski 1997, Brooks et al. 2000) added receivers at 15 -- 25~GHz 
(15-mm), 30 -- 50~GHz (7-mm) and 85 -- 105~GHz (3-mm); for details see Gough 
et al. (2004) and Moorey et al. (2008). A further extension of the highest 
frequency band to 77 -- 117~GHz (as already available on the 22-m Mopra 
antenna; Moorey et al. 2006) is desirable in the future. In 2010 the 1.5~GHz 
(20-cm) and 2.3~GHz (13-cm) bands were combined into one broad band covering 
the frequency range from 1.1 to 3.1~GHz (effectively 10 -- 30-cm, now referred 
to as the 16-cm band). See Table~\ref{tab:atcaprop} for a summary of some basic 
ATCA properties. Because the ATCA system temperature and primary beam FWHM
vary significantly across the available frequency range, these are shown in
Figs.~3 \& 4.

\begin{figure} 
 \mbox{\psfig{file=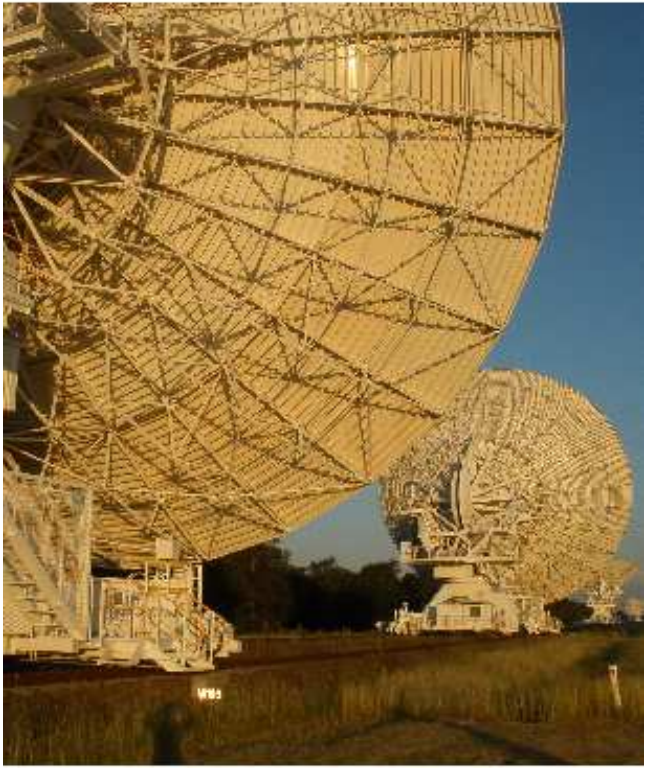,width=8.5cm,angle=0}} 
\caption{\label{atca2}
  The Australia Telescope Compact Array (ATCA). Displayed are several 
  antennas on the East-West track pointing at a source near the horizon. }
\end{figure}

\begin{figure*} 
 \mbox{\psfig{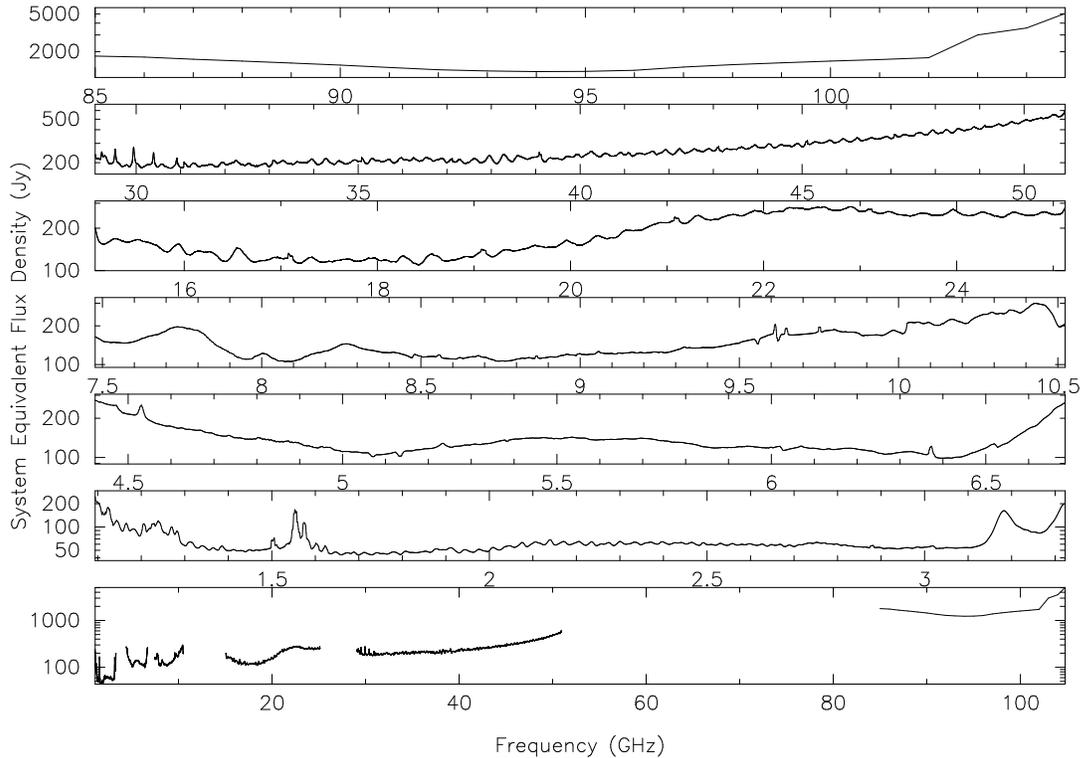}} 
\caption{\label{cabb-tsys}
  ATCA system equivalent flux density (in Jy, log scale), for each observing 
  band, obtained at high elevation and under reasonable observing conditions. 
  --- Note that one to four 2~GHz IF bands can be placed in each row, 
  separated by no more than 8~GHz, for simultaneous single or dual 
  polarization observations.  --- The measurements at frequencies from 1.1 
  to 50~GHz were made by Jamie~Stevens in Jul/Aug 2010 using the CABB system, 
  while the 3-mm measurements (80 to 105~GHz) were made by Tony~Wong in Sep 
  2004. The values are based on hot-cold load measurements and include 
  the atmosphere at the time of the observation. }
\end{figure*}

\begin{table*} 
\caption{Some ATCA properties. --- See the {\em ATCA Users Guide (Table~1.1)} 
   for more details and up-to-date information.}
\label{tab:atcaprop}
\begin{tabular}{lccccccc}
\hline
ATCA observing bands  & 16-cm$^{\star}$  & 6-cm & 3-cm & 15-mm & 7-mm & 3-mm \\
                      &  (L/S) &  (C) & (X)  & (K)   & (Q)  & (W)  \\
frequency range [GHz] & 1.1 -- 3.1 & 4.4 -- 6.7 & 7.5 -- 10.5
                      &  15 -- 25  &  30 -- 50  &  85 -- 105 \\
number of antennas    &  6 &  6 &  6 &  6 &  6 &  5 \\
number of baselines   & 15 & 15 & 15 & 15 & 15 & 10 \\
primary beam FWHM     & 44\arcmin -- 16\arcmin 
                      & 10\farcm7 -- 7\farcm4 
                      &  6\farcm3 -- 5\farcm1 
                      &  $\sim$2\arcmin 
                      &  $\sim$70\arcsec 
                      &  $\sim$30\arcsec \\
\hline
\end{tabular}

{\bf Notes:} 
   ATCA observing information can be found at
   {\em www.narrabri.atnf.csiro.au/observing}, including a link to the CABB 
   Sensitivity Calculator which is highly recommended to obtain observing 
   characteristics (e.g., \tsys) at specific frequencies and correlator 
   settings (see also Fig.~3). The ATCA primary beam size (in arcmin) can 
   be approximated by $50 / \nu$ where $\nu$ is the observing frequency in 
   GHz; the {\sc miriad} task {\sc pbplot} provides details of the primary 
   beam model (see Fig.~4).
   $^{\star}$ In 2010 the 1.5~GHz (20-cm) and 2.3~GHz (13-cm) bands were 
   combined into one broad band covering the frequency range from 1.1 to 
   3.1~GHz (now referred to as the 16-cm band). Note that the 3- and 6-cm 
   bands can be used simultaneously.
\end{table*}

\begin{figure} 
 \mbox{\psfig{file=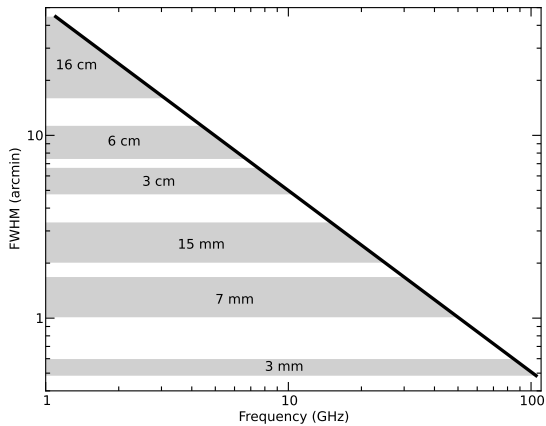,width=8.5cm,angle=0}} 
\caption{\label{ATCApbeam}
  ATCA primary beam FWHM as a function of frequency; values were computed 
  with the {\sc miriad} task {\sc pbplot}. The currently available frequency 
  coverage across all ATCA bands from 1.1 to 105~GHz range is indicated by
  the grey-shaded bands (see also the linear display of the ATCA frequency
  coverage in Fig.~3).} 
\end{figure}

In this paper, we describe the design and construction of the Compact Array 
Broadband Backend (CABB), the first observational results, and the scientific
potential of the upgraded instrument. The sections are as follows:
(2) CABB overview,
(3) CABB design,
(4) CABB installation and operations,
(5) CABB data reduction software, 
and (6) first CABB results. \\

The design and construction of other correlators for radio interferometers
such as Expanded Very Large Array (EVLA) and ASKAP are described in the 
respective overview papers by Perley et al. (2009) and DeBoer et al. (2009).
SKA correlator advances are discussed by Bunton (2005) and software 
correlators by Deller et al. (2011). 

\section{CABB Overview}\label{sec:cabboverview}  

The Compact Array Broadband Backend (CABB) upgrade, described here, has 
provided a new wide-bandwidth correlator for the ATCA, which is significantly 
more versatile and powerful than the original correlator (Wilson et al. 1992). 

The maximum bandwidth of the ATCA has been increased from 128~MHz to 2~GHz 
(dual polarization) in each of two independently tunable intermediate 
frequency (IF) bands, while CABB also increased the velocity resolution 
and delivers full Stokes parameters in all observing modes. Furthermore, 
the digitisation level has improved from 2-bits to 9-bits, increasing 
correlator efficiency and consequently lowering \tsys. In the following 
we list the advances provided by the new system.

\begin{itemize}
\item The correlator is supplied with an 8~GHz wide IF band from the front-end.
  Within this, the observer can tune (within the receiver limits, see Table~1) 
  two independent 2048~MHz windows (dual polarization) for correlation. This 
  represents a factor of at least 16 increase in the useable bandwidth over 
  the original ATCA correlator. Each 2~GHz window can be split into 2048, 512, 
  128, or 32 primary channels (see Table~2).
\item Up to 16 high velocity resolution {\em zoom} windows can be placed 
  anywhere within each 2048~MHz band. Each {\em zoom} window covers the width
  of one "continuum channel" in the primary band and further splits it into 
  2048 high-resolution spectral channels (as illustrated in Fig.~9). Basic 
  correlator configurations are listed in Table~2.
\item CABB combines high spectral resolution with full Stokes imaging, e.g.,
  needed for the study of Zeeman splitting in masers. All Stokes parameters 
  are computed by the new correlator in all available modes. The original ATCA 
  correlator was unable to provide cross-polarization correlations when used 
  in modes with high spectral resolution.
\item Sampling resolution and digitisation accuracy has increased from 2-bits 
  with the original correlator (Wilson et al. 1992) to 9-bits with CABB (see 
  \S~3.2), increasing the correlator efficiency from $<$0.88 to $\approx$1 
  (i.e., reducing the ATCA \tsys\ by $\sim$14\%), the dynamic range and 
  tolerance to radio frequency interference (RFI). 
\item Internally, CABB constructs a polyphase filter bank, providing spectral 
  channels that are largely independent with $\approx$80~dB isolation from 
  adjacent channels. This avoids the ringing that was commonly seen with the 
  original ATCA and other correlators while observing narrow spectral lines.
\item Modes providing high velocity resolution (for spectral line studies), 
  high time resolution (for the study of fast transients), or pulsar binning 
  come as an addition to the basic wide-bandwidth modes. 
\item CABB also provides auto-correlation data.
\end{itemize}

\begin{table} 
\centering
\begin{tabular}{lrr}
\hline
configuration & \multicolumn{2}{c}{channel width} \\
              & primary band & secondary band \\
\hline
CFB 1M--0.5k  &  1.0 MHz     &  0.488 kHz \\
CFB 4M--2k    &  4.0 MHz     &  1.953 kHz \\
CFB 16M--8k   & 16.0 MHz     &  7.812 kHz \\
CFB 64M--32k  & 64.0 MHz     & 31.250 kHz \\
\hline
\end{tabular}
\caption{Basic CABB configurations.}
\label{tab:cabbmodes}
\end{table}

\begin{figure*} 
 \mbox{\psfig{file=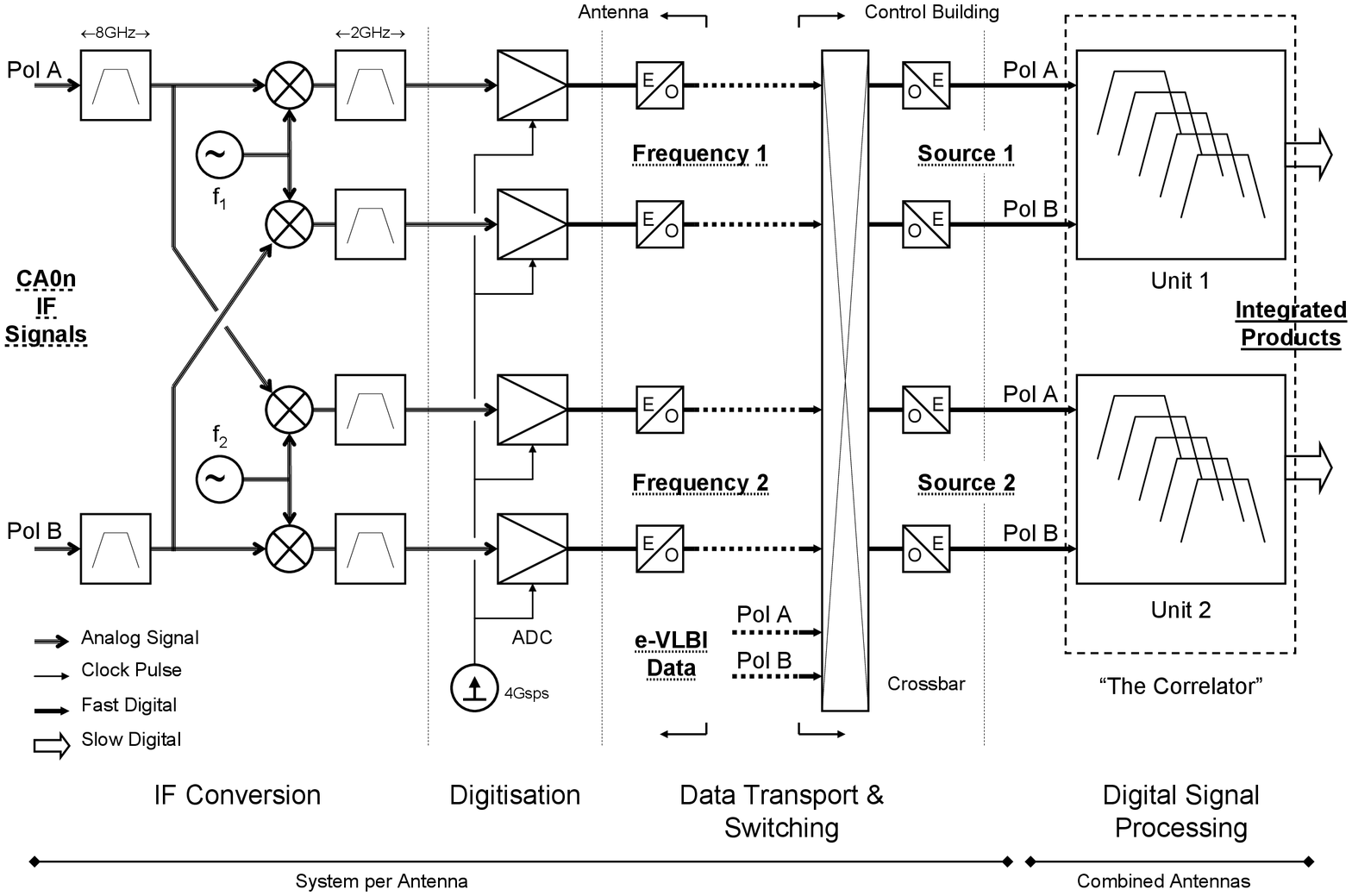,width=12cm,angle=0}}
\caption{\label{cabb1}
   CABB System Overview. }
\end{figure*}

These improvements have a major impact on the scientific ability of the ATCA
(see examples in \S~6), including the following:

\begin{itemize}
\item the much larger bandwidth reduces the time required to reach any 
  particular continuum sensitivity, and the increased sampling depth allows 
  for higher dynamic range and lower \tsys;
\item narrow, independent channels allow for precise excision of narrowband
  interference;
\item the division of the primary IF bands into a large number of channels 
  substantially improves the $uv$-coverage for any single observation when 
  multi-frequency synthesis is used (see \S~5.4);
\item at low frequencies (1 -- 10~GHz), CABB's large fractional bandwidth 
  makes it possible to study the spectral behaviour of continuum sources 
  without resorting to frequency switching (see \S~6.7);
\item at high frequencies (16 -- 105~GHz), CABB's large frequency coverage 
  allows projects to simultaneously observe multiple spectral lines that fall 
  within 8~GHz of each other; it also provides broad velocity coverage to carry
  out reliable searches for molecular lines at high redshifts (see \S~6.9);
\item the CABB {\em zoom} modes enable simultaneous observations of multiple 
  spectral lines at high velocity resolution (see \S~6.3);
\item the CABB {\em zoom} channels can also be concatenated to provide a wide 
  velocity range, while maintaining high velocities resolution (see \S~6.5);
\item the correlator can be configured to provide the most suitable compromise
  of sensitivity, speed and resolution for any particular observation. This 
  is achieved by varying the width of each channel in the primary bands (see 
  Table~2).
\end{itemize}

\section{CABB System Design} 

The two main objectives of the ATCA broadband upgrade were to increase the 
sensitivity and versatility of the instrument while providing a test-bed for 
technologies which were judged to be important in the development of the 
Square Kilometre Array (SKA). The improvement in sensitivity was achieved by 
increasing the maximum available bandwidth from 128~MHz to 2.048~GHz. A {\em
Field Programmable Gate Array} (FPGA) based correlator gave the required 
versatility, with its capability of being configured into many different 
operating modes. An important goal was to improve the spectral line capability 
of the instrument, particularly at the higher observing frequencies. 
Technologies employed, such as multi-bit digitising, high bandwidth digital 
data transfer over fibre optic cables, and advanced signal processing, were 
all considered to be applicable to the SKA.

The first stage of the CABB backend selects two independently tuneable 2~GHz 
bands from the active receivers. Two orthogonal linear polarizations are 
available from each band. The four resulting 2~GHz wide IF bands are digitised 
and sent over fibre optic cables to the central control building. Here they 
enter the correlator, where the online signal processing takes place. A block 
diagram of the system is shown in Fig.~\ref{cabb1}, and details of the signal 
path and the processing are described in the following sections.

Associated analogue-to-digital conversion (ADC) requirements are pushing the 
limits of high-speed sampling and quantisation techniques. Nine bit data 
samples and 17-bit (minimum) internal signal paths through to the integrators 
provide performance approaching analogue correlation but with the obvious 
advantages of digital processing. Combined with excellent channel selectivity 
the system is significantly more robust against strong interference than 2-bit 
correlators (Ferris \& Wilson 2002). 

These new technologies are direct candidates for signal processing on the 
proposed Square Kilometre Array (SKA) and its precursors, such as ASKAP, the
Australian SKA Pathfinder. CABB includes a total of eight antenna ports, 
allowing two antennas to be added to the existing six antennas of the ATCA, 
providing both a more powerful instrument and a mature test bed for the SKA 
pathfinders. \\

In the following we describe the IF conversion system (\S 3.1),
the digitisers and data transmitters (\S 3.2),
data transport and reception (\S 3.3),
the digital filter bank correlator (\S 3.4),
the signal processing (\S 3.5),
the correlator hardware implementation (\S 3.6),
the rear transmission module (\S 3.7),
the front board (\S 3.8),
and the control software (\S 3.9). 
In \S 3.10 we briefly look at the role of CABB in demonstrating novel 
technologies for SKA. The last sub-section (\S 3.11) gives an overview
of the CABB installation.


\begin{figure*} 
 \mbox{\psfig{file=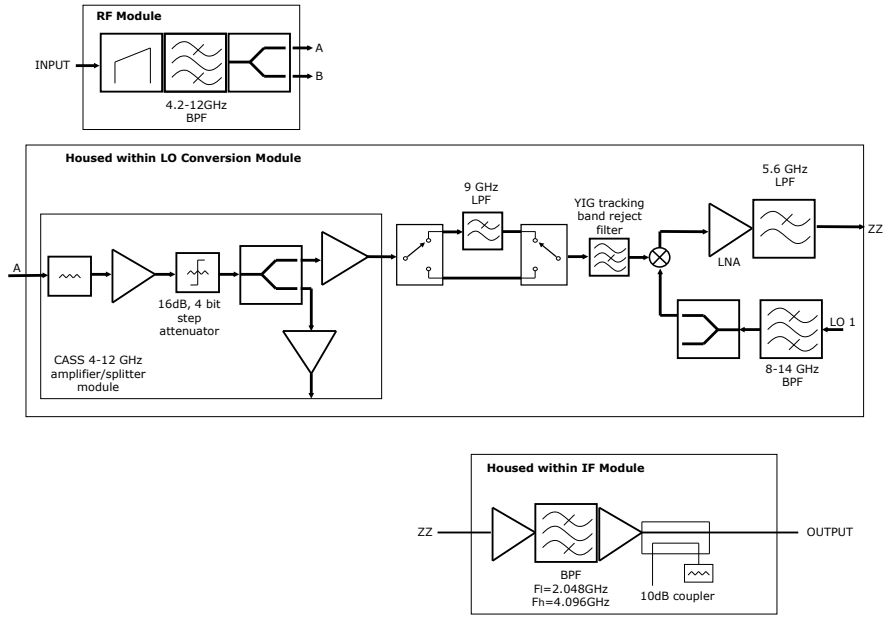,width=12cm,angle=0}}
\caption{\label{cabb2}
   The three sections of the CABB Conversion system: (1) radio frequency (RF) 
   module, (2) conversion module, (3) intermediate frequency (IF) module. 
   BPF = bandpass filter, LPF = lowpass filter, LNA = low noise amplifier.}
\end{figure*}

\subsection{IF conversion system} 

The modular CABB IF conversion system resides in a shielded rack at the 
vertex of each antenna of the ATCA. The functionality of this system (see 
Fig.~\ref{cabb2}) includes primary band filtering 4.2 to 12~GHz, stepped 
level control of 16~dB, sub-band selection for image management and frequency 
translation to suit the digitiser requirement. Definition of the IF band is 
achieved by anti-aliasing filters covering 2.048 to 4.096~GHz, corresponding 
to the second Nyquist zone of the analog to digital converter. 

The signal flow in Fig.~\ref{cabb2} indicates the fundamental elements, which 
were selected with a significant emphasis on low cost where possible and the 
reuse of ATNF designed assemblies. Each input pair (two polarizations) is  
passively split allowing for two discretely tuneable 2~GHz wide slices of 
the incoming primary signals. The assembly is physically divided into three 
sections, the radio frequency (RF) module, the conversion module, and the 
intermediate frequency (IF) module, briefly described below: 

\begin{itemize}
\item Primary band filtering, loss slope equalisation and passive power 
  splitting is achieved in the {\em RF module}. 
\item Slope compensated low noise amplification, digitally controlled step 
  attenuation, image suppressing sub-band selection, and frequency translation 
  are achieved in the signal processing part of the {\em conversion module}. 
  Support for maintenance is provided by low level monitoring outputs. 
  Suppression of potential local oscillator (LO) leakage contamination from 
  the split frequency design is provided by both fixed and tracking filtering. 
  The provision of a YIG (yttrium iron garnet) tracking LO rejection filter is 
  only made tractable by the integrated tuning management incorporated into 
  the optically slaved LO system (see below). 
\item The {\em IF module} provides slope compensated amplification, low level 
  monitoring outputs and anti-aliasing band filtering. 
\end{itemize}

In an effort to minimise the impact on system stability when moving to the 
increased fractional bandwidth required by CABB, a newly conceived and designed
group of filters following a castellated wall ridged waveguide topology (Bowen 
et al. 2010) were used. These filters have low insertion loss, low sensitivity 
to temperature change and high component repeatability. They are used in the 
primary input filtering, LO noise filtering and the IF anti-aliasing filter. 

Two central site reference frequencies, each in the range of 8 to 14~GHz, are 
transmitted to each antenna via separate intensity modulated optical fibre 
links. The antenna YIG oscillators are then offset phase locked to the optical 
references. The specific choice in YIG oscillators for the frequency conversion
was heavily biased toward oscillators having low in-band phase noise and very 
low broadband noise in order to limit the coherent noise contamination seen 
at the IF.

The system's dynamic range suffers as a result of the need to place a hard 
bound on the IF module's saturated output power level to prevent degradation
or destruction of the digitisers. The original low-end specification of the 
primary band was modified from 4 to 4.2~GHz due to excessive direct leakage 
into the output band. Triple balanced mixers were chosen due to their enhanced 
leakage suppression and lower spur product generation to further reduce 
spurious signal generation.  

\begin{figure} 
 \mbox{\psfig{file=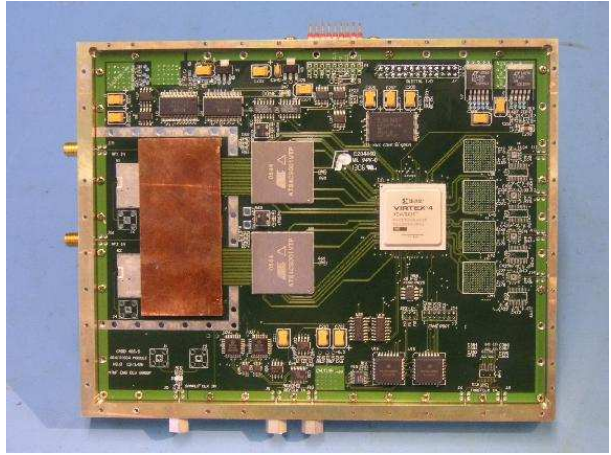,width=8cm,angle=0}} 
\caption{\label{cabb-adc}
     The CABB analogue-to-digital converter (ADC) / data transmitter 
     module; see \S~3.2. }
\end{figure}

\subsection{Digitisers and Optical Data Transmission} 

A major innovation required to realise the CABB specifications was the 
development of an analogue-to-digital converter (ADC; see Fig.~\ref{cabb-adc}) 
capable of supporting the 4.096~Gigasamples/s (GS/s) rate required for 
processing the 2.048~GHz of bandwidth demanded by the upgrade. The project 
ambitiously aimed to implement quantising at bit levels significantly higher 
than those traditionally undertaken in radio astronomy correlators. Typically, 
two bit sampling has been used. The major goal was to allow for faithful 
handling and removal of RFI in the signal processing as well as increased 
sensitivity. There were no commercially available ADC devices that met these 
specifications at the time the design commenced. 

The need for high dynamic range, mostly due to RFI (but also required for 
deep fields or in the vicinity of extremely strong radio sources) makes it 
necessary to use sampling at high bit levels. In the noise dominated case 
(i.e., classical astronomical observation), the loss due to 2-bit (4-level) 
quantisation is large (Van Vleck \& Middleton 1966; Cooper 1970), leading to
a significant ($>$14\%) increase of the system temperature, \tsys. A high 
resolution digitiser with sufficient bits can be operated with negligible
increase of \tsys, i.e. the correlator efficiency is $\approx$1.


The solution was to develop an ADC subsystem composed of two commercial 
2~GS/s 10-bit ADC chips operated in an interleaved fashion. The same input is 
distributed to the two converters with their sampling times in precise 
anti-phase, to give an equivalent 4.096~GS/s 10-bit converter. This requires 
extremely tight matching of the two channels in terms of sampling phase, gain, 
bandwidth, and DC offset. High resolution calibration and trimming circuitry 
that can control these parameters to the level required has been incorporated 
on the board, and algorithms developed to automatically measure, and adjust, 
these trims in real time. Experience has shown that the system stability is
such that only a single calibration is required. The board and ancillary 
systems are housed in a module that is located in each antenna's shielded rack.

Performance achieved over the 2.048 to 4.096~GHz CABB IF band at full sampling 
rate is a signal-to-noise ratio of typically 38~dB, equivalent to approximately
six effective digitiser bits, and spurious free dynamic range of 39~dB. 

In addition to analogue-to-digital conversion the CABB ADC/Data Transmitter 
module performs framing and transmission of the ADC data. It takes the 
40~Gigabits/second (Gbits/s) of data from the digitiser and frames it into 
four 10~Gbits/s serial data streams which are passed to laser driver chips 
driving externally modulated lasers at four different optical wavelengths.  
These are wavelength division multiplexed onto a single fibre for transport 
from the antenna to the central control building. Of the ten bits of sampled 
data, nine bits are transported over the link. The remaining link overhead is 
used for implementing a custom synchronisation and timing protocol that 
enables each of the 10~Gbits/s data streams to be re-aligned and allows 
precise tracking of the delays on each path back to the control building. 

The digitiser/transmitter module has been designed to be as general as possible 
to allow it to be used in other applications apart from the CABB upgrade. It 
has, for example, been used as two independent 1~GHz bandwidth ADCs in the 
ATNF Pulsar Digital Filterbank systems at the Parkes 64-m telescope (Manchester
et al. 2010), for radio transient detection experiments\footnote{During the 
  CABB development phase (before March 2009), three antennas of the ATCA were 
  used to hunt for ultra-high energy (UHE) neutrinos using the lunar Cherenkov
  technique. For details of the LUNASKA experiments see James et al. (2010) 
  and McFadden et al. (2008). By-passing the original ATCA IF system, the 
  teams obtained 600~MHz of bandwidth in the then available 20-cm band (1.2 
  to 1.8~GHz). Each polarised data stream was fed through an analogue
  de-dispersion filter (Roberts \& Town 1995) before being sampled by the 
  CABB ADC board at 2~GS/s with 8-bit effective precision.}
with both Parkes and the ATCA and as an RFI transient detector at the ASKAP 
Boolardy site.

\begin{figure} 
 \mbox{\psfig{file=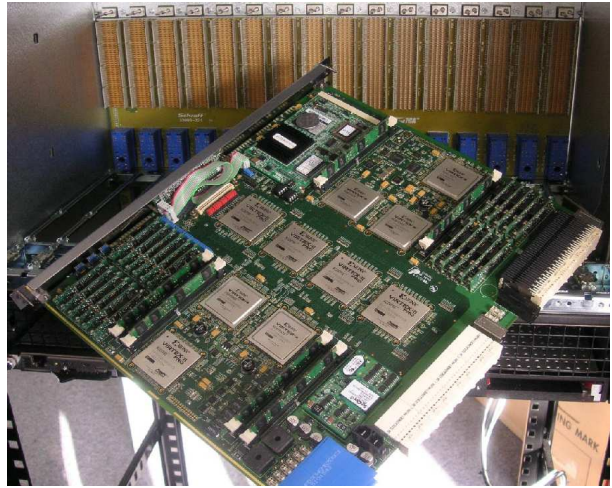,width=8cm,angle=0}} 
\caption{\label{cabb-dsp}
   The CABB Digital Signal Processing (DSP) board. }
\end{figure}

\subsection{Data Transport and Reception} 

The digitised data are transported to the correlator in the central control 
building via a single optical fibre feed from each digitiser. Each optical 
fibre carries four 'colours' at the 200~GHz ITU frequencies of 193.8~THz, 
193.6~THz, 193.4~THz and 193.2~THz. Each laser is co-packaged with an 
electro-absorption modulator and is intensity modulated at 10 GBits/s.

In the central control building, the four colours on each fibre are 
de-multiplexed and each optical signal is then detected by a Receive Optical 
Sub-Assembly (ROSA) which incorporates a PIN detector and a transimpedance 
amplifier. The electrical outputs to the correlator are Current Mode Logic 
differential signals. The ROSAs are part of the Rear Transition Module (RTM),
which forms the interface between the fibre transmission and the correlator. 
The RTM is described in more detail below.

\begin{figure*} 
 \mbox{\psfig{file=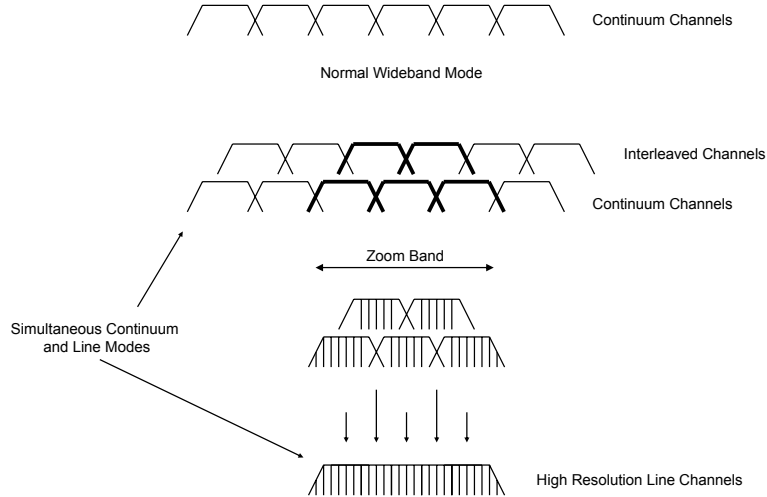,width=12cm,angle=0}}
\caption{\label{cabb3}
  CABB provides up to 16 {\em zoom} windows which can be allocated
  individually or in groups. Here we show an example where five {\em zoom} 
  windows are concatenated (overlapped in steps of 0.5~channels) to increase 
  frequency coverage while maintening high velocity resolution and high
  dynamic range. }
\end{figure*}

\subsection{The Digital Filter Bank Correlator} 

The central feature of CABB is the FX-type spectroscopic correlator. Its 
design supports primary channelisation with 1, 4, 16 and 64~MHz resolution 
consistent with science requirements across the ATCA input frequency range 
of 1.1 to 105~GHz. For each frequency channel on each baseline the four 
cross-products of the two polarizations are formed. Also, for each frequency 
channel and for each antenna the two polarization auto-correlations and the 
cross-polarization product are formed, including a facility for synchronous 
extraction of a switched calibration signal. The latter is injected coherently 
into the orthogonal linear polarised receivers for polarization calibration.
All products are always available across the full 2.048~GHz signal bandwidth. 

At the same time up to 16 {\em zoom} windows (channels) at the primary 
resolution bandwidth, but half its channel spacing, may be selected for 
secondary processing into 2048 channels each. Again, all four polarization 
products are recorded. The available configurations are shown in Table~2. 
When two or more adjacent {\em zoom} windows are chosen, they are aggregated 
to provide a seamless high resolution spectrum across their net bandwidth. 
Passband ripple from the primary filter bank is automatically removed. 
This system provides for the simultaneous observation of multiple widely 
spaced spectral lines, plus the underlying continuum. An example of a 
{\em zoom} band setup is shown in Fig.~\ref{cabb3}.

Antenna signals with the primary channelisation may also be formed into 
beams for VLBI, pulsars, spacecraft tracking, etc., with the native linear 
polarizations added in quadrature to provide circular polarizations as 
required.

There are two such correlator units, one for each of the ATCA's independently 
tuneable observing frequencies. Where the front-end bandwidth allows, 
observations at full specifications across a total of 4~GHz are possible.  
Alternatively the units may be tuned to the same input frequency to repeat 
the observation with a different set of correlator configuration parameters.  

\begin{figure*} 
 \mbox{\psfig{file=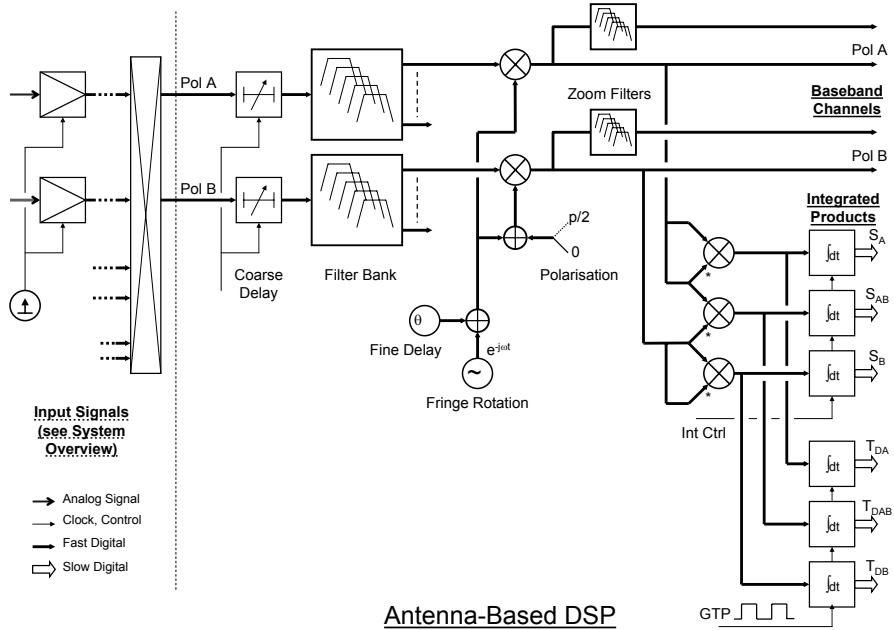,width=12cm,angle=0}} 
\caption{\label{cabb5}
    CABB Signals from orthogonal linear polarizations are digitised, 
    channelized, fringe-stopped and calibrated. }
\end{figure*}

\subsection{Signal Processing} 

\begin{figure*} 
\begin{tabular}{cc}
 \mbox{\psfig{file=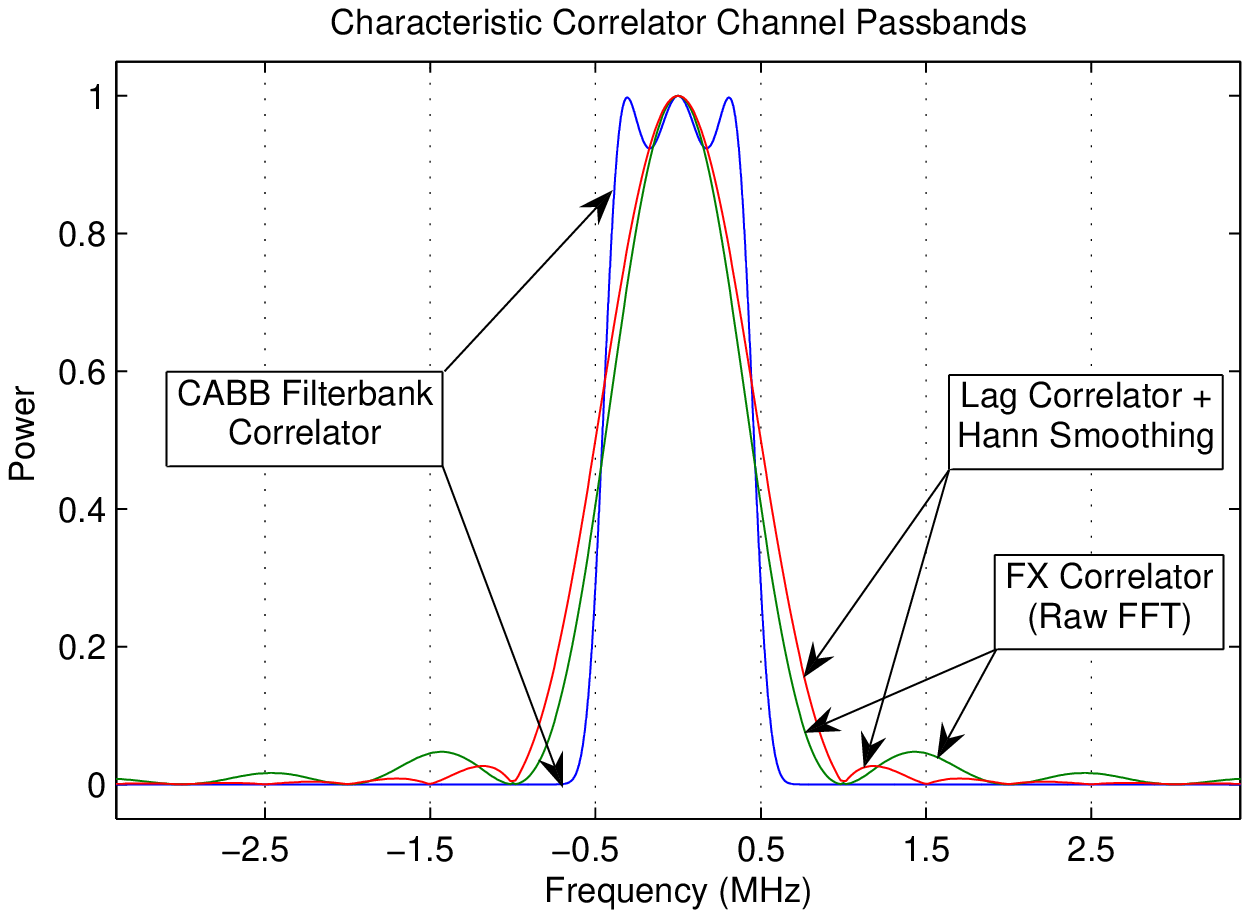,width=9.0cm,angle=0}} &
 \mbox{\psfig{file=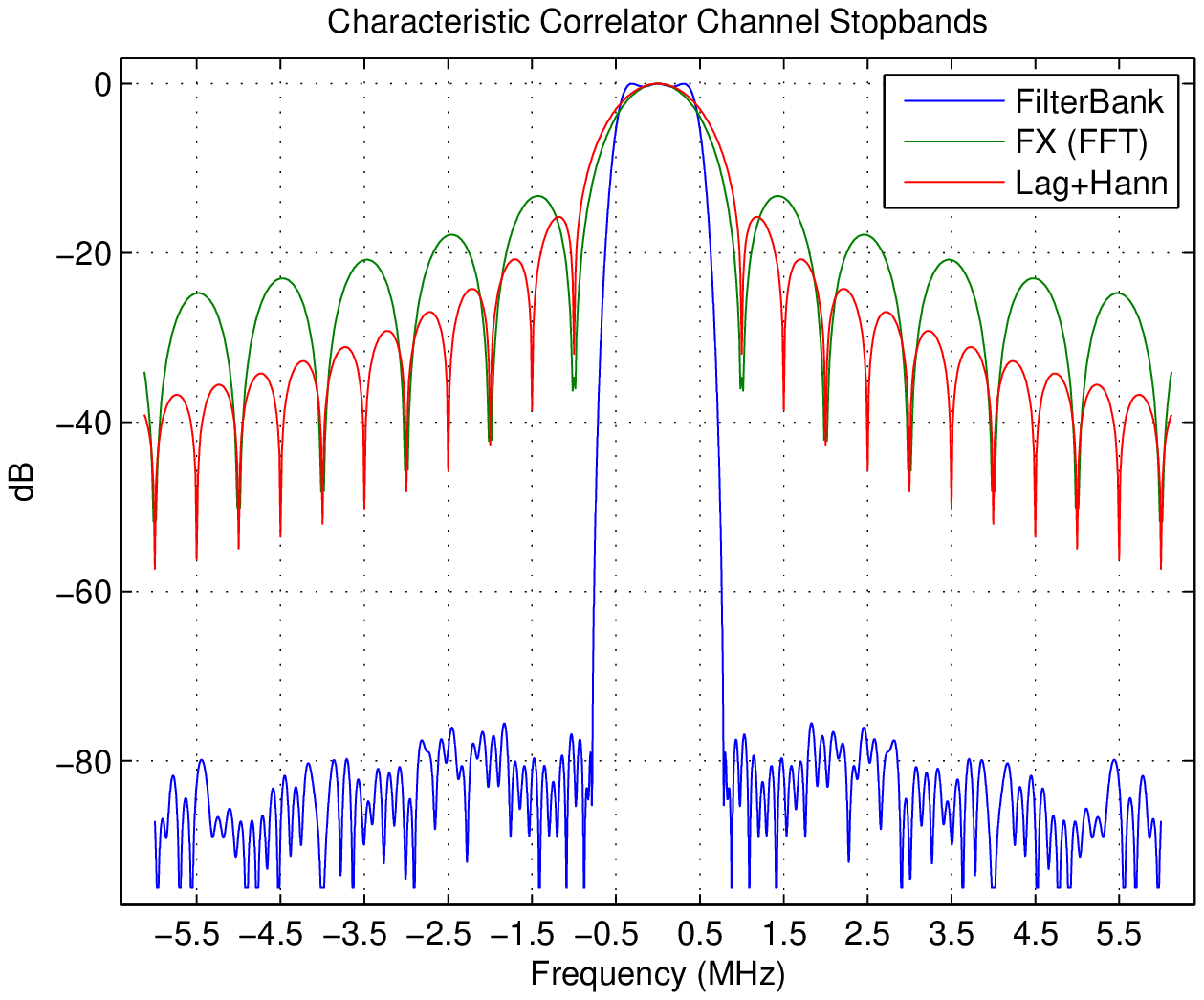,width=7.8cm,angle=0}} \\
\end{tabular}
\caption{\label{corr}
  Comparison of the CABB Filterbank 1~MHz channel 'square' response (blue 
  lines) with those of classic 'FX' correlators (green lines) and lag/'XF' 
  correlators (red lines); the FX response function is sinc$^2$($\nu$), while 
  the lag response is sinc($\nu$) where $\nu$ is the frequency. The absence 
  of sidelobes in the CABB Filterbank channels is remarkable; for details 
  see \S~4.2. {\bf Left:} The three functions are plotted on a linear scale 
  over $\pm$3.5~MHz and, {\bf (right)}, on a log scale over $\pm$6~MHz.}
\end{figure*}

\begin{figure*} 
 \mbox{\psfig{file=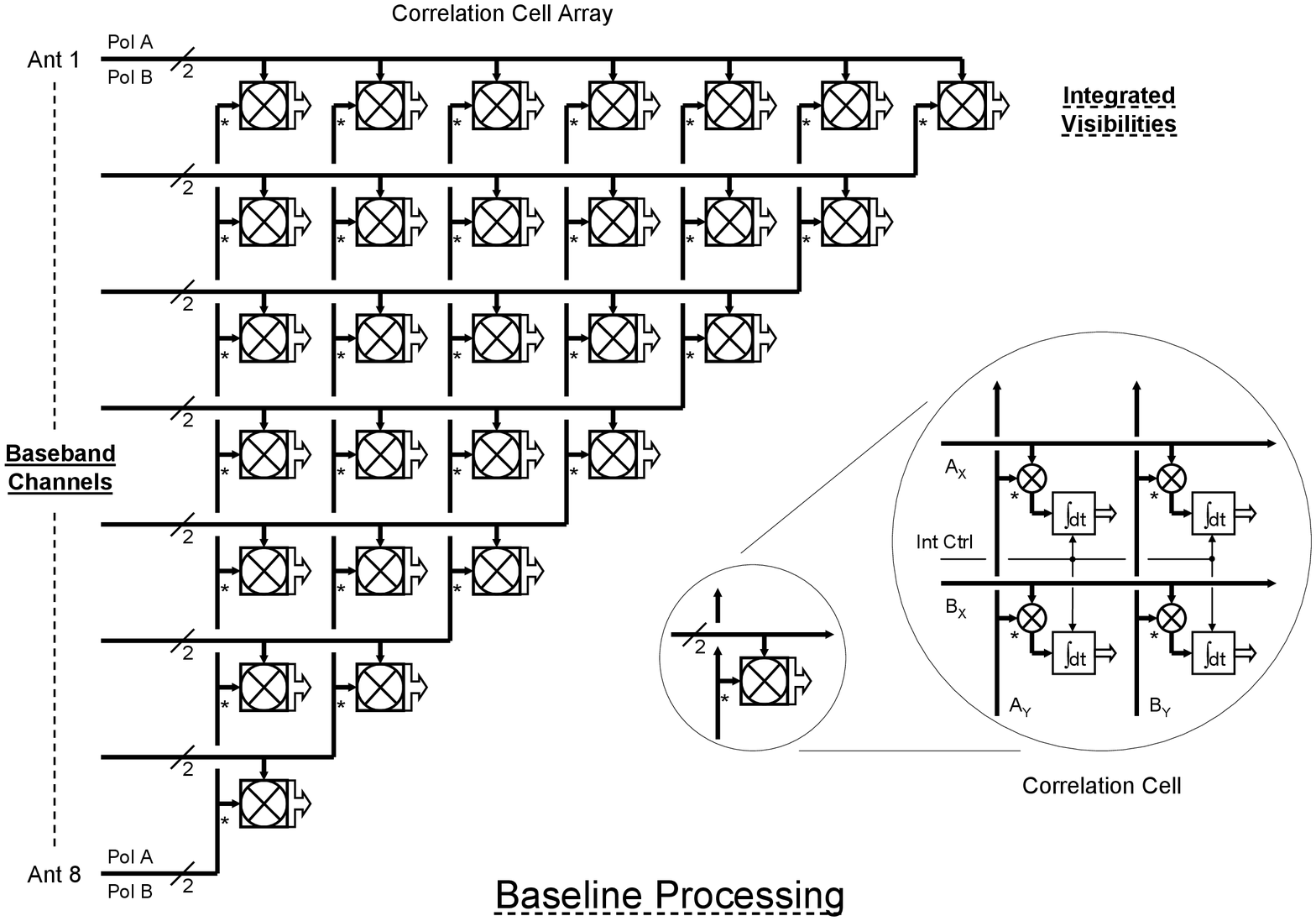,width=12cm,angle=0}} 
\caption{\label{cabb6}
  CABB Polarisation pairs from each antenna are multiplied with the
  conjugated pairs from every other antenna. All four cross products 
  are formed and integrated.}
\end{figure*}

The essential structure of the correlator signal path is shown in 
Figs.~\ref{cabb5} \& \ref{cabb6}. Beginning with high resolution IF data from 
the ADCs, the bulk of the geometric delay is removed to the nearest sample 
time in the `Coarse Delay' units. The 2~GHz wide signals are then analysed 
into many baseband channels as per Table~2 in the primary polyphase digital 
filter banks (PDFBs). The flat-topped, sharp cut-off, deep ($\sim -80$~dB) 
stopband filter shape achieved (see Fig.~\ref{corr}) far exceeds the 
requirements of good measurement. Its purpose is to prevent the spread of 
strong RFI from channel to channel. The filter banks output only the positive 
frequency signal components to make subsequent processing easier.

Fringe rates are stopped on a per channel basis. Since the channel bandwidths 
are much smaller than their centre frequency on the sky, the timing residuals 
left by the `Coarse Delay' are manifest as all but constant phase shifts. This 
is accurately removed by adding an offset, `Fine Delay', to the fringe phase.  
An additional $\pi$/2 can be added to the B polarization phase whenever 
necessary to generate circularly polarised beams in the beam formers.

After the `Fringe Rotators' the signals are wavefront-aligned at all 
frequencies across all antennas. At this point (see Fig.~\ref{cabb5}) the 
user-selected subset of primary channels is fed to narrow band PDFBs to 
produce the high resolution {\em zoom} channels. The antenna auto-correlations 
and X/Y phases\footnote{Instrumental phase difference between X and Y
  polarization channels}
are also formed, using synchronous integration to extract the signature of 
the low-level switched noise diode calibration signal which is injected into 
the receiver horn. These signals are used to monitor and refine system 
calibration.

Baseband channels from all antennas are brought together in the `Correlation 
Cell Array' (see Figure~12). In each cell the signals from one antenna are 
multiplied by the conjugate of the corresponding signals from another antenna, 
and the products integrated in simple accumulators. All four possible products 
are formed. Thus each cell computes the complex visibilities for one baseline. 
These are normally unloaded at ten second intervals, but shorter integration 
times are supported (110ms in pulsar binning mode and 20ms in high time 
resolution). {\em Zoom} channel processing follows the same form and is 
synchronised with the primary channels so that all output data have the same 
epoch.

Signals from all antennas can be added to produce phased array beams, with the 
option of applying complex weights to steer the beam around within the primary 
antenna pattern. The resulting signals are available outside the correlator 
for processing in other equipment.

The digital signal path begins with 9-bit ADC data, and word size expands with 
natural growth plus the addition of guard bits. There is no clipping or
re-quantisation, and noise addition due to rounding and aliasing is carefully
controlled. The dynamic range of the ADC is preserved so that expected
variations in system temperature, front-end passband slope and ripple, and
strong astronomical sources are accommodated without resort to gain control 
or nonlinear correlation corrections, and there is still headroom to tolerate 
RFI without distortion or compression.

The combination of high dynamic range in the time domain and high selectivity
in the filter bank (see Fig.~\ref{corr}) ensure robustness against RFI. This 
linear, stationary signal path will provide an ideal platform for future RFI 
removal instigations (Ekers \& Bell 2001).

Contrary to tradition there is no `Correlator Chip' ASIC in this design. 
Instead all signal processing takes place in FPGAs wherein ready access to a
very high density of logic resources allowed the development of an instrument 
with a level of sophistication unattainable in a hard-wired circuit. In 
addition, flexibility through reprogrammability has supported progressive 
development with evaluation on telescopes, multiple independent configurations 
(listed in Table~2), a shift from one generation of FPGAs to the next to take 
advantage of the latest technology, and the spinoff of several related but 
distinct instruments. Separating the signal processing from the hardware also 
allowed the parallel and largely independent development of the latter, a 
significant engineering project in its own right. A by-product of this 
flexibility has been the complexity of modes which are possible, leading 
to an increase in time needed for testing and commissioning.

\subsection{Correlator Hardware Implementation} 

\begin{figure*} 
 \mbox{\psfig{file=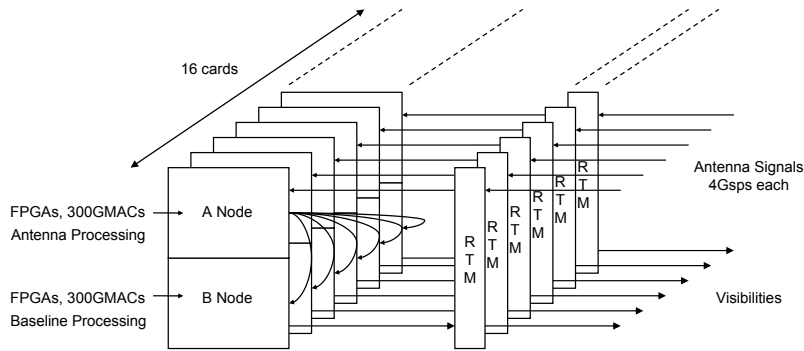,width=12cm,angle=0}}
\caption{\label{cabb7}
    Schematic of the Advanced Telecom Compute Architecture (AdvTCA) chassis
    (see \S~3.6). Every A~node connects to every B~node with four 2.5~Gbps 
    links. The Rear Transition Module (RTM) provides the physical interfaces
    to the external environment. }
\end{figure*}

The architecture of the hardware platform, as depicted in Fig.~\ref{cabb7}, is 
driven by the need to connect signal paths from the antenna-based processing 
to the baseline-based processing, the details of which have been described in 
the previous section. It is well known that for large interferometers this 
apparently mundane task (which grows as the square of the number of antennas) 
is the main obstacle and becomes the major cost driver. For the modestly sized 
CABB correlator the solution was to adopt the newly announced Advanced Telecom
Compute Architecture (AdvTCA) chassis, a commercial item developed by and for
the telecommunications industry. Its backplane includes a mesh network, 
providing a high bandwidth link from every slot to every other slot, exactly 
as required. Each of 16 card slots can accommodate a large `Front Board' which 
connects to the mesh and supports the signal processing, and a smaller directly
linked board, the Rear Transition Module (RTM), which provides the physical 
interfaces to the external environment. 

The CABB Front Board has half of its FPGA resources connected to the input side
of the mesh, supporting most of the antenna-based digital signal processing 
(DSP; see Fig.~\ref{cabb-dsp}) for one IF (polarization) for one antenna. The 
remaining resources are driven by the output side of the mesh and support the 
baseline-based DSP, plus the antenna-based Integrated Products since they 
require both polarizations. Thus each mesh input node sinks the full spectrum 
from 1/16 of the Frequency~1 (or 2) antenna IFs, and each mesh output node 
sources 1/16 of the spectrum from the full set of Frequency~1 (or 2) antenna 
IFs. In this way one AdvTCA chassis supports a correlator for eight antennas, 
the six ATCA dishes plus two further antennas, providing a mature test bed for 
the SKA pathfinders.

\subsection{The Rear Transition Module} 

The functions of the Rear Transition Module (RTM) are:
\begin{itemize}
\item to receive the four optically transported 10~Gbit/s data signals from 
  one digitiser and to convert them back to electrical signals using one ROSA 
  per 'colour' and then synchronise and merge these four separate data streams
  into one parallel data stream;
\item to receive and buffer the main system clock signals of 256~MHz, 128~MHz,
  32~MHz and 160~MHz;
\item to receive and buffer supplementary control signals, including a one
  pulse per second timing reference signal;
\item to provide four, four-lane Infiniband channels, with each lane capable 
  of up to 3.125~Gbit/s data rate, for connection to or from external systems
  as required;
\item to provide a mass storage device for the Front Board's embedded 
  processor; and
\item to provide two ethernet connections to the control and data networks.
\end{itemize}

Most of the logic functionality is achieved using one FPGA. This FPGA may be
set to be configured from an onboard PROM however it is generally configured
from the Front Board's embedded processor.

\subsection{The Front Board} 

The complexity of the `Front Board', also known as the CABB Signal Processing
card, is indicated by its specifications. These include a 26-layer printed 
circuit board (PCB) measuring $322 \times 280 \times 3$mm-standard AdvTCA form 
factor, 4188 components, and $\approx$20000 nets \& 230000 holes.
The major elements in the processing card are:
\begin{itemize}
\item Ten signal processing FPGAs that have the following functions: 
   (i) to perform coarse data delay functions and implement a filter bank; 
   (ii) to provide $4 \times 16$ = 64 high speed 2.5~Gbit/s communications 
   channels, four to each of the other fifteen cards and one loopback. Each 
   communications channel is full duplex, meaning that there is an independent 
   10~Gbit/s TX and 10~Gbit/s RX data path to every processing card from every 
   other card; (iii) to implement a correlator; and (iv) provide for an 
   external high speed data interface for applications such as VLBI and 
   spacecraft tracking.
\item A small control computer with a Geode LX800 processor and 128~Mbytes of 
  memory and onboard 10/100 Mbit/s Ethernet. The system communicates with the 
  processing FPGAs via a standard PCI bus. An additional Ethernet interface 
  chip is installed on the processing card to enable 1~Gbit/s Ethernet 
  communications.
\item Two control FPGAs, one distributing control and monitoring data to all
  signal processing FPGAs and the other distributing accurate timing signals
  to all FPGAs. Both interface directly to the above system over the local 
  PCI interface bus.
\item Six mini dual in-line memory modules (DIMMs) of DDR2 RAM, each supporting 
  up to 1~GByte. Each correlator FPGA is connected to two modules and the delay
  FPGA is connected to two modules.
\item A flexible clock distribution network allowing either externally provided
  reference signals from the RTM, or locally generated signals, to be
  distributed to all processing FPGAs. Each processing card contains four
  local oscillators operating at 66~MHz, 200~MHz, 266~MHz and 156.25~MHz. The
  clock distribution network can be configured to shutdown if a temperature
  sensor surpasses a preset threshold.
\item System health monitor points for all onboard-generated voltages, all 
  FPGA temperatures and PCB temperatures around the board.
\item An isolated power conversion module with a 48~V input generating all 
  the required voltages for the FPGAs onboard. A large distributed heat sink
  covers all signal processing FPGAs to distribute heat over the entire card
  and reduce hot spots. Each card is rated at 200~W.
\end{itemize}

All signal processing FPGAs are connected to each other with parallel low 
voltage differential signalling (LVDS) data busses of varying widths. All 
data pairs run at a common data rate of 512~Mbps. Data arrives from the rear 
IO card across the AdvTCA zone~3 connector and is received by the coarse delay 
FPGA. It is then transferred to the two filter bank FPGAs. These FPGAs are
interconnected allowing the signal processing required for the filter bank
to be split across both. Each filter bank FPGA is connected to one of the
data distribution FPGAs. Data is distributed between these four FPGAs as
well as to and from every other processing card in the AdvTCA rack. Every 
data distribution FPGA has a dedicated link to each of the two correlator
FPGAs, and each correlator FPGA has access to two modules of DDR memory for
data accumulation. 

\begin{figure} 
 \mbox{\psfig{file=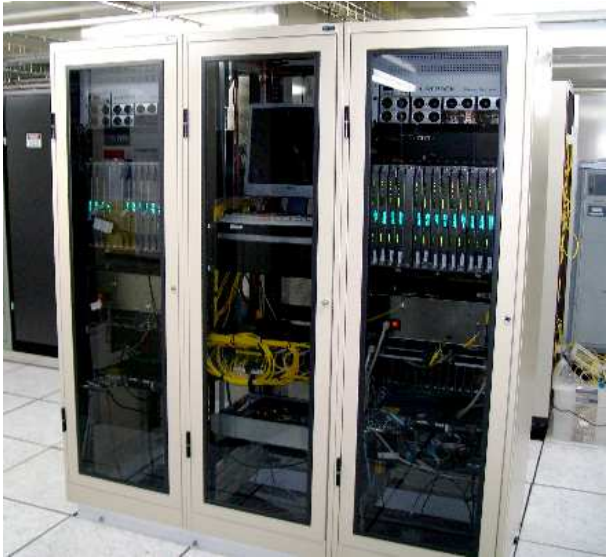,width=8cm,angle=0}}
\caption{\label{cabb-rack}
  The CABB rack as installed in the fully shielded ATCA correlator room.}
\end{figure}

\subsection{Control Software} 

Control of the CABB hardware and firmware has been integrated into the standard 
correlator control software package, which itself has been developed over many 
years to support backend systems at all ATNF observatories. The major new 
functions required for CABB were antenna hardware switching control, local 
oscillator setting, correlator configuration control, delay tracking, 
procedures for the calibration of delay, phase and amplitude, system monitoring
and data collection and archiving. The fact that much of the functionality of 
the previous backend system has been moved into the correlator required a 
significant expansion of the correlator control software. Examples of this 
are the delay tracking and narrow band conversion systems, which were 
previously located in special purpose hardware in the antennas. 

The delay tracking software uses the GSFC Mk-5 VLBI Analysis Software CALC 
(Gordon 2004). A custom interface to the CALC package has been created to 
facilitate its use in the real-time CABB application.

\subsection{Square Kilometre Array (SKA) Development Role} 

The construction of CABB (see Fig.~\ref{cabb-rack}) embodied the role of 
demonstrating enabling
technologies for the SKA. Novel features include the digital filter bank
correlator architecture, the high 2~GHz signal bandwidth and 4~GHz total
bandwidth, high resolution digitisation followed by a linear, stationary,
signal path, RFI robustness, and simultaneous continuum and spectral line
observations at full specifications across the full bandwidth. The correlator
provides extra ports to incorporate two additional external antennas.
Wideband data produced by CABB provide challenging input to a new generation
of off-line data reduction and imaging software tools under development for
ASKAP and the SKA. \\

The Mopra wide-band system, the CABB 2~GHz correlator in Narrabri, the SKA
Molonglo Prototype (SKAMP) correlator at Captain's Flat (Adams, Bunton \&
Kesteven 2004), the Murchison Widefield Array (MWA) correlator (Lonsdale et
al. 2009) and the Parkes Testbed, where a 48 (upgradable to 192) input 
beamformer has been implemented, can be considered prototype systems for the 
design of the ASKAP correlator (DeBoer et al. 2009).

\section{CABB Installation and Operations} 

The installation of CABB required a significant shutdown of the Compact Array 
totalling six weeks in March/April 2009. The task was lessened somewhat through
progressive installation over the year of an interim CABB system. This interim 
system, which ultimately provided a single 2~GHz bandwidth, with dual 
polarizations for five of the six ATCA antennas, was able to be operated in 
parallel with the original correlator system, allowing valuable comparisons 
and cross-checks to be made. In concert with 
the signal transmission and processing hardware changes, modules to interface 
the current suite of receivers to the new system have been fabricated and 
installed. CABB operations and scientific commissioning were interleaved with 
engineering maintenance periods in late April 2009. The new CABB system has 
also required significant changes to the array control and monitoring software, 
which has taken place in parallel with the hardware changes. In addition, the 
ATCA data analysis package {\sc miriad} has required revision and extensions 
to enable the significantly larger CABB data files to be processed.

The first mode available for general observing consisted of $2 \times 2048$~MHz
bandwidths with 2048 channels across each band, corresponding to a 1~MHz 
spectral resolution. For VLBI and NASA tracking a single 64~MHz channel 
observing mode was also made available. Table~2 summarises the basic modes;
currently available are:

\begin{itemize}
\item CFB~1M--0.5k: a bandwidth of 2~GHz divided into $2048 \times 1$-MHz 
      channels and (optionally) a fine resolution of 0.5~kHz in up to four 
      {\em zoom} bands in each of the two IF bands has been available since 
      May 2010. --- The first observations with $16 \times 1$~MHz {\em zoom} 
      bands were obtained on 17 Dec 2010 (see Fig.~\ref{sagb2zoom16}).
\item CFB~64M--32k: a bandwidth of 2~GHz divided into $32 \times 64$-MHz 
      channels and (optionally) a fine resolution of 32~kHz in up to four 
      {\em zoom} bands in each of the two IF bands. Since October 2010 a 
      single 64~MHz channel (divided into 2048 sub-channels) is available 
      in each IF. We hope that all 32 channels and {\em zoom} channels will 
      be available later in 2011.
\item {\em pulsar binning modes} were successfully tested in December 2010
      (see \S~6.13). The minimum time bin is $\sim$110$\mu$s, allowing, for 
      example, 32 phase bins across the period of a 3.5 milli-second pulsar.
\item {\em high time resolution modes} ($\sim$10ms) was implemented in January
      2011, with 4~MHz frequency resolution across each of the two 2~GHz IF
      bands.
\end{itemize}

\begin{figure} 
 \mbox{\psfig{file=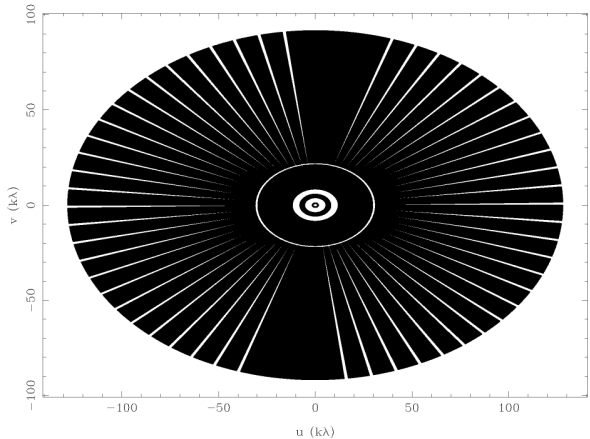,width=8.5cm,angle=0}} 
\caption{\label{uvcoverage}
    The spectacular $uv$-coverage obtained with CABB 2~GHz bandwidth in 
    a full 12-h synthesis observation. Shown here are projected baseline 
    tracks obtained during an observation of Pictor\,A (\S~5.3) at 6-cm 
    with the ATCA in the 6A configuration. }
\end{figure}

\subsection{CABB Milestones} 

The CABB project commenced in January 2002 and had at its core a new digital
signal processing (DSP) system based on a novel polyphase Digital Filter
Bank (PDFB) structure developed at ATNF.  
Key milestones on the path to completion of CABB included provision of the 
Mopra Spectrometer (MOPS\footnote{Previously developed ATNF 8~GS/s 2-bit 
  sampler chips, as used in the Mopra Spectrometer (MOPS) were unsuitable 
  for the CABB upgrade. Two bit sampler systems have significant quantization 
  errors, increasing the noise and causing distortions. While they have 
  adequate RFI rejection at high frequencies, they are unsuitable for the 
  stronger RFI often encountered at lower frequencies. MOPS delivers a maximum 
  bandwidth of 8~GHz, split into four overlapping IFs of 2.2~GHz each. Three 
  modes are provided, giving resolutions of 2.2~MHz, 270~kHz or 34~kHz.};
see, e.g., Walsh et al. 2010, Urquhart et al. 2010, Muller et al. 2010) 
in 2006, delivering 8~GHz of bandwidth, and the Pulsar Digital Filterbanks 
for the 64-m Parkes dish (Ferris \& Saunders 2004).  

CABB replaces the Compact Array's original signal processing and digital 
correlator system (Wilson 1992). Those systems, at the time of their design 
and construction in the 1980s, were "state-of-the-art" and gave the ATCA a 
competitive edge when it began operating in 1990 (Frater \& Brooks 1992). 

First fringes with the interim CABB system were obtained on 23 May 2008 
between CA02 and CA03. On 12 Aug 2008, a three-baseline system was first 
demonstrated with the addition of CA05, with fringes obtained on a single
polarization with a 128~MHz bandwidth, enabling further testing to be carried 
out. In early November 2008, full 2~GHz auto-correlation spectra were being 
obtained from CA01 to CA05, followed by the first CABB image (single IF, 
128~MHz bandwidth, full polarization) in early December. Comparison with data 
from the original (pre-CABB) correlator indicated the CABB system was working 
well. In late December 2008 and early January 2009, the CABB 
bandwidth was gradually increased toward the full 2~GHz, with additional 
imaging observations being made to test the flow of data from the correlator 
and through {\sc miriad}. In February 2009, the first observations utilising 
the full 2048 channels were made with the interim CABB system (five antennas, 
single IF). On 23 Feb, CA06 was taken off-line and preparations started for 
the full CABB installation. A six-week shutdown ran from 2 Mar to 14 Apr 2009, 
followed by a week of scientific commissioning (in the H168 array). Scheduled 
observing started on 22 Apr 2009. 

The CABB shutdown diary, available on-line, provides a lively documentation of
the progress during that period. CABB fringes were progressively obtained to 
the Compact Array antennas, with first CABB fringes to all six antennas 
obtained on 1 Apr 2009. Upgrades to the ATCA operations and monitoring 
software and to the popular {\sc miriad} data reduction package were carried 
out in parallel (see Section~5). \\

\begin{figure} 
 \mbox{\psfig{file=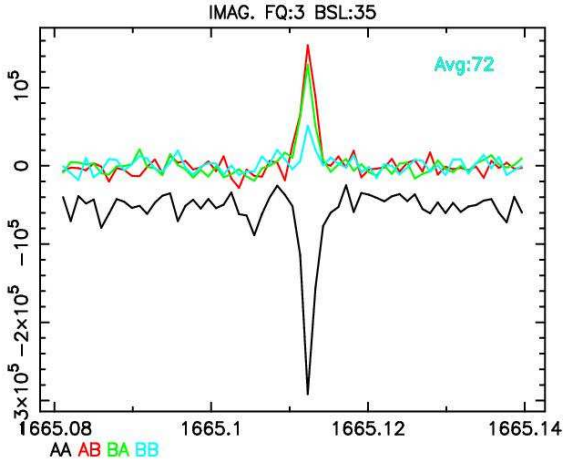,width=8cm,angle=0}} 
\caption{\label{cabb-zoom}
   Progress towards {\em zoom} modes: a single baseline partial {\em zoom} 
   spectrum of the maser OH~213.705 from ATCA CABB observations on 4 Mar 
   2010. Intensity (arbitrary units) is plotted against frequency in MHz.
   AA, AB, BA and BB are the four polarisation products between antennas 
   CA03 and CA05, showing the imaginary part of the cross correlation so 
   as to highlight the high degree of polarization.}
\end{figure}

First scientific results, obtained by operations and astrophysics staff as 
well as ATNF co-supervised students, were presented during the CABB Science 
Day on 13 May 2009 (available on-line). Initially, only bands from 3-mm to 
6-cm were available for testing, with poor weather conditions severely limiting
observations in the 3-, 7- and 12-mm bands. The 13- and 20-cm bands (now 
replaced by a combined and much broadened 16-cm band) were available for 
testing from 15 May 2009. The large number of "dead channels" (due to missing 
CABB boards in April and May 2009) made it difficult to obtain presentable, 
wide-band spectra. \\

The first CABB zoom mode observations with a single ATCA baseline were taken 
in March 2010 (see Fig.~\ref{cabb-zoom}). In December 2010, using all ATCA
baselines, we successfully demonstrated spectral line observations with $16 
\times 1$~MHz {\em zoom} channels providing very high velocity resolution, 
and also pulsar binning mode. High time resolution modes are currently being 
tested. In Section~6 we present a selection of results obtained during the 
CABB commissioning week in April 2009 and afterwards. 

\subsection{The primary 2~GHz spectrum} 

A spectrum over the full bandwidth of 2~GHz is always observed and recorded 
(strictly, 2048~MHz, since all bandwidths and frequency separations are 
precisely $2^{\rm n} \times 1$~MHz) for both IF bands. All four polarization 
products are computed, along with the auto-correlations and cross-correlations 
(interferometer output) for each of the six antennas of the ATCA. The chosen 
IF bands must be entirely contained within 8~GHz of each other (see Figure~3, 
for an overview of the currently available bands). Note that the 2~GHz band 
edges are defined by an analogue filter (as were the bands of the original 
ATCA correlator), so that \tsys\ increases gradually (by less than a factor
two) towards the outermost edges of the band. 

For this basic 2~GHz spectrum the maximum resolution is 1~MHz. Each channel 
has a 'square' response, and thus there are 2048 independent channels across 
the full spectrum (see Fig.~\ref{corr}. This means that Hann (or "Hanning")
smoothing, which is often used to remove `ringing', is not required. Internal 
to the correlator, an output of 4096 channels is routinely computed (i.e., 
with additional channels interleaved with a half-channel shift) so as to 
allow flexibility in positioning of {\em zoom} bands (see Figs.~\ref{cabb3} 
\& \ref{cabb-zoom-schematic}). To enable the set-up of {\em zoom} bands with 
various widths, the following observing modes are planned for the primary 
band: $2048 \times 1$~MHz, $512 \times 4$~MHz, $128 \times 16$~MHz and $32 
\times 64$~MHz (see Table~\ref{tab:cabbmodes}).

In Dec 2010 the original 20- and 13-cm band receivers were upgraded to a single 
10--30-cm (or nominally 16-cm) band, providing an instantaneous frequency 
coverage from 1.1 to 3.1~GHz (although the usable bandwidth is reduced by RFI).
The new receivers have an improved sensitivity over the original 20- and 13-cm 
receivers, and include new ortho-mode transducers, significantly improving the 
polarization performance towards the top end of the band. 

In the 6- and 3-cm bands, which are generally observed simultaneously, the 
existing front-ends allow a 2~GHz bandwidth to be used for each band (see 
Fig.~3).

In the 15-mm (previously referred to as 12-mm), 7-mm and 3-mm bands, two IF
bands may be selected within an 8~GHz bandwidth. In the 7-mm band, both band 
centres must be either greater than 41~GHz (the point at which the conversion 
changes from lower side-band to upper side-band) or both less than 41~GHz.
For example, methanol lines at 36 and 44~GHz cannot be observed at the same 
time.

\subsection{RFI and bad channels} 

Single-channel interference spikes are found one-quarter, one-half and 
three-quarters of the way across the 2~GHz band and are due to the way the 
interleaved ADC samples are generated (see \S~3.2). When using 2048 channels, 
these are located in channels 513, 1025~(the centre channel) and 1537. 
Because they are always contaminated with self-generated interference, the 
channels are flagged by the correlator before the data are written to disk.

Further interference spikes  are commonly observed due to harmonics of the 
640~MHz data clock and the 4096~MHz ADC clock. These occur in channels 129, 
257, 641, 769, 897, 1153, 1281, 1409, 1793 and 1921 (in 2048 channel mode). 
The correlator does not by default flag these channels (but can be instructed 
to) because they are not always correlated.

\subsection{CABB cycle time} 

The CABB correlator defaults to writing out one set of spectra every ten
seconds. If required, this interval can be decreased to as little as one
second. Without {\em zoom} windows, the correlator will output 3.9 MByte 
per cycle, which corresponds to a data rate of 1.4 GByte per hour for ten 
second cycles. With a full set of zoom windows this rate may increase 
nearly twenty-fold.

\begin{figure} 
 \mbox{\psfig{file=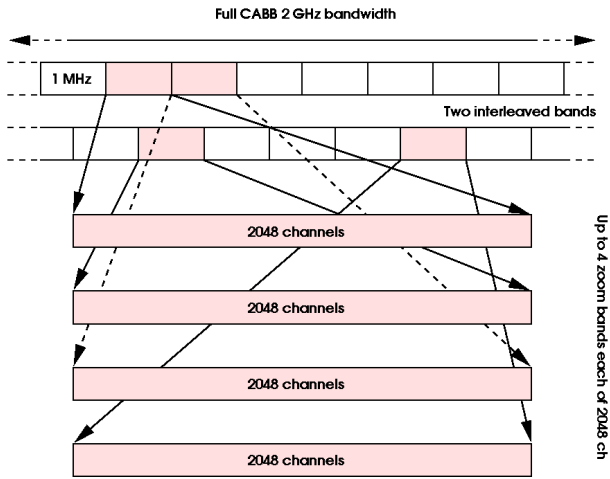,width=8cm,angle=0}} 
\caption{\label{cabb-zoom-schematic}
  Illustration of a $4 \times 1$~MHz CABB {\em zoom} mode setting. 
  CABB internally computes two interleaved bands, each with 2048 channels, 
  and up to 16 {\em zoom} windows may be selected (in addition to the 2048 
  channels across the 2~GHz primary band). The interleaved channels, with 
  half-channel offsets, allow flexibility in the positioning of {\em 
  zoom} bands so that spectral lines do not fall at a channel edge.}
\end{figure} 

\subsection{Zoom modes} 

CABB provides up to 16 {\em zoom} windows in each of the two selected IF 
bands; these are in addition to the full 2~GHz primary spectrum. The four 
basic correlator configurations are listed in Table~2. 

In each zoom band, the output will always be 2048 independent non-overlapping 
channels.  Interleaved channels are not computed, but the channels will still 
be clean 'square' channels, with negligible spillage (i.e., no ringing from 
narrow lines). The choice of bandwidths for the {\em zoom} windows is 1, 4, 
16, and 64~MHz. Each of the 16 {\em zoom} windows must be the same width as 
the "continuum channels" selected for the primary 2~GHz spectrum. The channel 
width of the primary 2~GHz spectrum can be chosen separately for each IF band.

The first full baseline ATCA observations with a CABB {\em zoom} mode were 
carried out in May 2010. Using $8 \times 1$~MHz {\em zoom} windows we were
able to observe eight methanol maser transitions around 25~GHz (see \S~6.3). 
This was followed by OH maser observations in July 2010, with CABB providing 
full Stokes polarization and high spectral resolution (Caswell \& Green 2011).
The first observations with $16 \times 1$~MHz {\em zoom} windows were obtained 
on 17 Dec 2010 (see \S~6.5).

The separation of selected {\em zoom} windows within the 2~GHz band can be 
any integral multiple of half the zoom bandwidth. To seamlessly cover a 
selected band, {\em zoom} windows need to be interleaved by a half width 
(see Figs.~\ref{cabb3} \& \ref{cabb-zoom-schematic}). For example, four 
interleaved 1~MHz channels cover 2.5~MHz of bandwidth or 16 interleaved 
1~MHz channels cover 8.5~MHz of bandwidth (see \S~6.5). Consolidation and
normalisation of interleaved channels takes place in the correlator, before
the data are written to disk.

\subsection{CABB scheduling} 

The addition of up to 16 {\em zoom} windows for each of two IF bands required 
an upgrade of the ATCA scheduling software. The old scheduler, {\em atcasched},
a terminal-based schedule creator, could not easily cope with another 32 input 
parameters. It has been replaced with a web-based scheduler called the {\em 
CABBScheduler}. The new scheduler runs as javascript in a browser and was 
written in java using the {\em Google Web Toolkit}.
It makes use of tabs and stacking to reduce the amount of information 
presented at once and integrates existing web-based tools like the ATCA 
calibrator database and the velocity calculator into the scheduler interface. 
Behind the scenes it uses server validation of the schedule timing using the 
same code as the observing program to ensure accurate scheduling. The 
scheduler calculates the {\em zoom} band channels based on the date, source 
position, velocity and line rest frequency. It also searches for calibrators 
near the position of interest and displays their characteristics such that
an informed selection can be made. While the scheduler is aimed at making it 
easy to manually create small schedules, there are facilities for bulk import 
of user created source catalogs and global search and replace of values. The 
schedule files are in a simple ASCII format with incremental scan to scan 
changes, making it relatively straightforward to produce large schedules 
with user written scripts.
   
Comprehensive ATCA observing information can be found at
{\em www.narrabri.atnf.csiro.au/observing}, including a link to the CABB 
Sensitivity Calculator which is highly recommended to obtain observing 
characteristics (e.g., r.m.s. noise per channel) at specific frequencies and 
correlator settings. 

As the implementation of {\em zoom} modes proceeds and our experience with 
CABB observing grows, the {\em CABBScheduler} is likely to accumulate further
options. Most recently, an option to set the channel range for calibration
purposes (in particular, delays and \tsys) was added, with the goal to avoid
RFI plagued channels in the respective observing bands.

After a proprietary period of 18 months, ATCA data are publicly available via 
the Australia Telescope On-line Archive (ATOA) at {\em atoa.atnf.csiro.au}. 

\subsection{CABB observing} 

The CABB correlator has radically altered the way the telescope operates. 
Extensive changes have been made to much of the control and monitoring 
software used by the ATCA, and a summary of these changes is presented here.

The coarse and fine attenuators have been replaced with CABB internal 
attenuators. The regular mm attenuators remain however, and progress is 
being made towards having specific attenuators for each cm receiver package 
as well. These attenuators continue to be controllable through {\sc CAOBS}.
The ATCA monitoring package, {\sc MoniCA} has been revised to include all 
the new CABB monitoring points, and can now also be used to control many of 
the devices it is attached to.
The primary control interface, {\sc CAOBS}, has remained much the same as for 
the original correlator, but there are a number of changes. The most 
important change has seen the introduction of separate commands for delay, 
phase and amplitude calibration at the beginning of an observing run, 
replacing the automated {\em cacal} procedure.

A variant of the mosaic observing mode available on the ATCA has been
introduced. Observations in the mosaic mode can cover a large contiguous
area of sky with numerous pointing centres. The new variant, called
``on-the-fly'' (OTF) mosaicing, reduces the overheads of moving between
pointing centres. For OTF mosaicing, the telescope slews continuously,
moving the beam by half its width during each correlator cycle. This mode
is made attractive with the broad bandwidths provided by CABB, as the
integration time can be reduced without incurring the usual slew-settling
overheads. In this way the time for each pass of the whole mosaic field
is reduced, improving the sampling of the $uv$-plane.

\begin{figure} 
 \mbox{\psfig{file=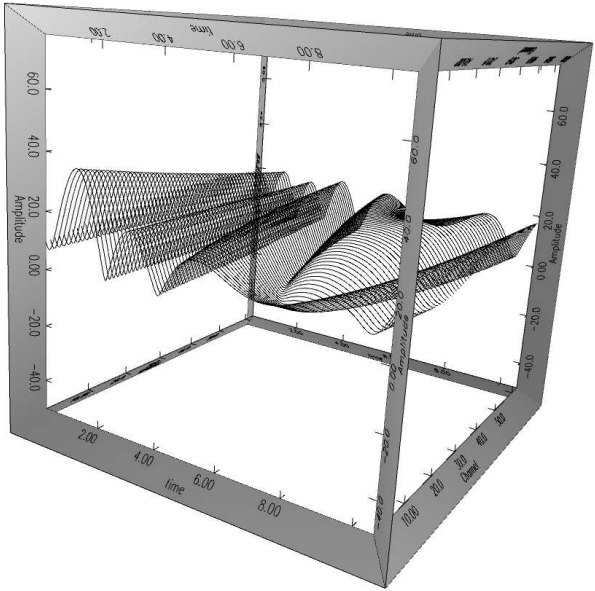,width=8cm,angle=0}} 
\caption{\label{bandpass}
  Visibility amplitudes varying over time and frequency for CABB observations 
  of the nearby FR-II type radio galaxy Pictor\,A (\S~5.3). The plot 
  is generated using the interactive tool {\sc plotvis} and shows visibility 
  amplitudes for a single baseline (CA02 -- CA03) of a 10-h observation with 
  a single 2~GHz bandwidth between 4.5 and 6.5 GHz, averaged into $64 \times 
  32$~MHz channels.}
\end{figure}

\section{CABB Data Reduction} 

The calibration, imaging and analysis of ATCA data is typically done in 
{\sc miriad}. As {\sc miriad} continues to be actively developed, it is
essential that the latest software updates are used to reduce CABB data. 

For continuum observations a new calibration strategy has been developed 
which ensures that amplitude calibration dependence on frequency is 
accounted for, regardless of the fractional bandwidth covered by the data. 
All ATCA users are encouraged to consult the {\sc miriad} Users Guide for 
the full details of this calibration strategy.

To improve spectral line data calibration, {\sc miriad} now generates 
time-variable bandpass solutions that have been found to reduce bandpass 
calibration errors in the 2~GHz band from 1\% to as little as 0.1\%. The 
{\em zoom} bands themselves have a digitally-defined bandpass that is 
corrected for online, and usually require no further correction during 
off-line reduction.  For gain calibration purposes, {\sc miriad} treats 
the zoom bands in the same way as the primary bands.

The greater instantaneous continuum sensitivity of CABB makes it easier to 
find suitable phase calibrators (close to the target source) and flux
calibrators over the full frequency range. For frequencies between 1~GHz 
and 25~GHz, the preferred ATCA flux calibrator is PKS\,1934--638. It has a 
known, stable flux, no detected linear polarization and only $\sim$0.01\% 
circular polarization (Reynolds 1994, Sault 2003). At high frequencies 
($>$50~GHz), the preferred flux calibrator is the planet Uranus. Its flux 
is known to vary with time, but it does so in a way that is understood and 
can be modelled (Orton et al. 1986, Kramer et al. 2008).
In the 7-mm band (30 -- 50~GHz) either PKS\,1934--638 or Uranus may be used, 
depending on the telescope configuration (in particular baseline length), 
frequency setup, and source elevation. Recent CABB measurements of 
PKS\,1934--638 (and Uranus) in the 7-mm band flux indicate 0.52 (0.70) Jy 
at 33~GHz and 0.35 (1.17) Jy at 45~GHz.

At frequencies below 10~GHz the increased bandwidth also means a larger number 
of channels will be affected by RFI, and extensive flagging of the data is 
required before processing. The flagging capabilities of {\sc miriad} have been
improved in an attempt to deal with this (see \S~5.1). The large fractional
bandwidth has to be considered for data reduction below $\sim$10~GHz. One 
approach is to split the data into several narrow frequency bands, then 
calibrate and image these separately. Correcting the images for the primary 
beam response, which varies with frequency (see Fig.~4), is then possible for 
each sub-band, before combining them to reach the full sensitivity, if 
required. Alternately, for single-pointing continuum observations at 4--10~GHz,
a 2~GHz wide band may be processed in one go as long as the multi-frequency 
synthesis (mfs) is used for imaging (options {\em mfs} and {\em sdb} in the 
task {\sc invert}) and deconvolution (task {\sc mfclean}). For observations 
that require high dynamic range and/or accurate polarization calibration the 
'divide and conquer' approach is still expected to yield better results as 
it can deal with variations in the calibration parameters with frequency. \\

A wiki for the {\em ATCA Users Guide} (atug.wikidot.com) and an {\em ATCA 
discussion forum} (currently at atcaforum.freeforums.org) have been set up 
for observers to discuss matters relating to CABB observing and data reduction.

Several tasks in the {\sc miriad} software package (Sault et al. 1995) have 
been modified and some new tasks have been added. This work will continue over 
the next few years. It is therefore important to (a) reduce data in the latest
version of {\sc miriad}, (b) read all relevant documentation, (c) check each 
step, each task, and intermediate results carefully, and (d) provide feedback 
to the development team ({\em email miriad@atnf.csiro.au}).

\subsection{Modified {\sc miriad} tasks} 

Some of the notable changes to {\sc miriad} tasks are:

\begin{itemize}
\item The option {\em rfiflag} in the task {\sc atlod} flags known, persistent 
  RFI, mostly seen in the low frequency range. The RFI details are regularly
  updated and stored in ascii table which can be downloaded and modified for 
  use in this task. The {\em birdie} option was updated to flag internal CABB 
  RFI which is fixed in channel number. Both options, {\em rfiflag} and {\em 
  birdie}, are recommended when loading CABB data as they eliminate a large 
  fraction of the RFI and faciliate subsequent flagging and data calibration.
\item A {\em maxwidth} parameter was added to the task {\sc uvsplit} to limit 
  the bandwidth of the output files, allowing  the data to be processed in 
  smaller frequency chunks.
\item The flagging program {\sc blflag} can now display data from all channels 
  separately using the {\em nofqav} option.
\item Time variability in the bandpass can now be solved for with the task
  {\sc mfcal}. The routines that apply calibration have been updated to 
  interpolate between the bandpass solutions.
\item The task {\sc mfboot} can now correct the spectral index across the 
  spectrum as well as the flux level when using planets for flux calibration 
  (typically used for mm observations).
\item Other tasks, in particular the plotting tasks, received minor updates 
  to cope with larger data volumes, larger numbers of simultaneous frequency 
  bands and much higher frequency resolution compared to that provided by the 
  original ATCA correlator.
\end{itemize}

\subsection{New {\sc miriad} tasks} 

\begin{figure} 
 \mbox{\psfig{file=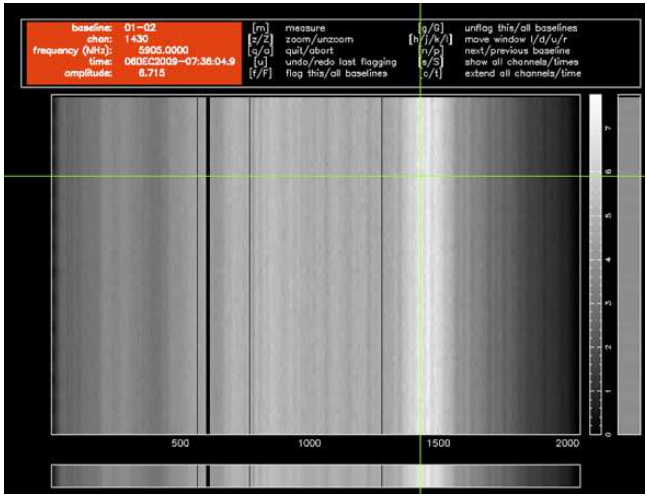,width=8.5cm,angle=0}} 
\caption{\label{pgflag}
   Screen-shot of the new {\sc pgflag} task in the {\sc miriad} software
   package (see \S~5.2).}
\end{figure}

Several new tasks have been written by the {\sc miriad} development team, 
including:

\begin{itemize}
\item {\sc pgflag} --- a pgplot based data editing tool (see Fig.~\ref{pgflag})
  replacing the task {\sc tvflag}, an excellent, interactive flagging task 
  which was designed for 8-bit displays;
\item {\sc mirflag} --- a task which allows automated flagging of a $uv$-data 
  set (specifically continuum data; see also {\sc tvclip} with options {\em
  notv}). It scans the data on a baseline-dependent and channel-dependent 
  basis to find outliers (either in amplitude or based on r.m.s. noise) and 
  flags these; it can be instructed to flag small or large clusters of data 
  points. Care should be taken to not accidentally flag astrophysically 
  important spectral lines.
\item {\sc plotvis} ---  provides an interactive, three dimensional view of 
  the visibilities (see Fig.~\ref{bandpass}) 
\end{itemize}

\subsection{Example: The Radio Galaxy Pictor\,A} 

\begin{figure} 
 \mbox{\psfig{file=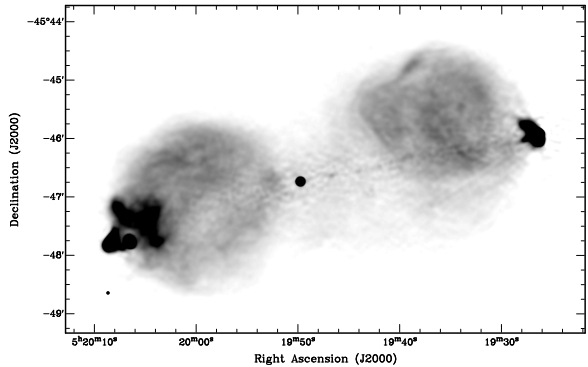,width=8cm,angle=0}} 
\caption{\label{pictora}
  ATCA 6-cm radio continuum image of the double-lobe radio galaxy Pictor\,A.
  made with CABB using 2~GHz of bandwidth. In this image, made using 
  observations in the 6A array (17 Jun 2009), 6D array (30 Aug 2009) and 
  EW352 array (06 Dec 2009) Lenc et al. achieved a dynamic range of 38,000:1
  and an angular resolution of 2\farcs2. Pictor A's complex structure, i.e. 
  its extended lobes, multiple hot-spots, narrow jet and bright core, present
  many imaging challenges (for more details see \S~5.3). }
\end{figure}

\begin{figure}
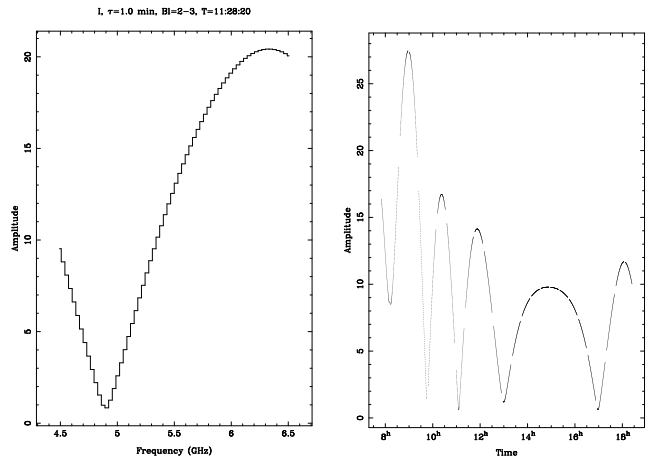
 
\begin{tabular}{cc}
 \mbox{\psfig{file=Emil.FigureX3a.ps,width=4cm,angle=0}}  &
 \mbox{\psfig{file=Emil.FigureX3b.ps,width=4cm,angle=0}} 
\end{tabular}
\caption{\label{pictora2}
  {\bf Left:} Spectral plot of Pictor\,A amplitude visibilities at a given 
  time for a single ATCA baseline (CA02 -- CA03) of a 10-h observation using 
  CABB with a single 2~GHz band between 4.5 and 6.5~GHz, averaged into $64 
  \times 32$~MHz channels. {\bf Right:} Plot of amplitude vs time for the 
  same observation.}
\end{figure}

The large bandwidth (divided into a large number of channels) now available 
with CABB provides a hugely improved $uv$-coverage for any single observation 
when multi-frequency synthesis (mfs) is utilised (see Fig.~\ref{uvcoverage}).
The almost-complete coverage reduces the need, in many instances, to observe 
the target with multiple ATCA configurations and thus alleviates issues 
associated with 
source variability and calibration differences between observations. However, 
care must be taken when imaging large or complex sources and sources with 
significant spectral variation. Large and complex sources, such as the nearby 
FR-II type galaxy Pictor\,A (Fig.~\ref{pictora}), introduce large variations 
in complex visibilities not only in time but also in frequency. 
Figures~\ref{pictora2}a \& b show the spectral plot at a given time and the 
amplitude visibility plot versus time. The time-frequency variability of the 
visibility amplitudes can more easily be seen using the new {\sc plotvis} tool 
(see Fig.~\ref{bandpass}) which provides an interactive, three dimensional 
view of the visibilities. Spectral differences between the various components 
of the source, i.e. the lobes, hot-spots, jet and the core, further complicate 
imaging and necessitate the use of multi-frequency deconvolution algorithms 
such as {\sc mfclean}.  

\section{CABB Results} 

CABB operations and scientific commissioning interleaved with engineering and 
maintenance blocks were carried out between 10--21 April 2009. Here we report 
on some of the results obtained during this period as well as more recent
observations.

\begin{figure} 
 \mbox{\psfig{file=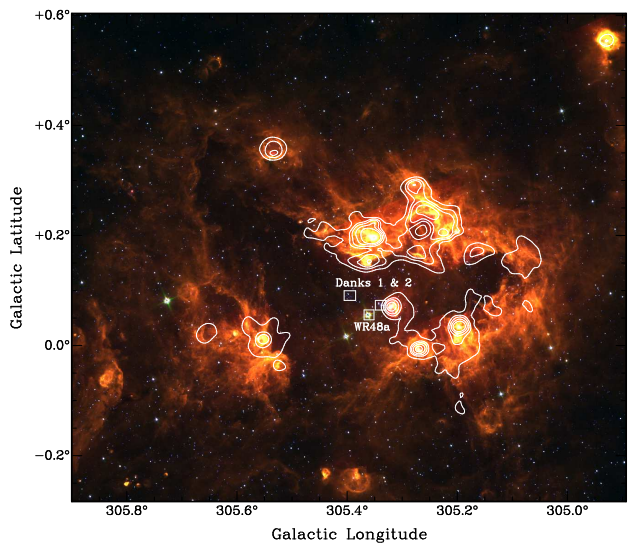,width=7.5cm,angle=-90}} 
 \vspace{0.5cm}
\caption{\label{G305_midIR}
  ATCA 6-cm radio continuum emission (contours) of the giant \HII\ region 
  complex G305 overlaid onto a three-colour Spitzer GLIMPSE image (using 
  the IRAC 4.5, 5.8 and 8.0\,$\mu$m bands coloured blue, green and red, 
  respectively, obtained from the Galactic Plane survey by Benjamin et 
  al. 2003). The continuum emission traces the distribution of ionized gas 
  associated with the \HII\ region complex.}
\end{figure}

\subsection{Star formation in the G305 \HII\ region}  

The G305 star forming region is one of the most luminous \HII\ regions in the 
Galaxy, centred on the open clusters Danks~1 \& 2 and the Wolf-Rayet star 
WR\,48a (Clark \& Porter 2004). This complex, which extends over roughly one 
square degree, is located at $l,b$ = 305\degr, 0\degr\ and at a heliocentric 
distance of $\sim$4~kpc, which places it in the Scutum Crux arm. The central 
clusters have ages of about 3--5~Myr and are located in the centre of a large 
\HII\ region, surrounded by molecular gas and dust. 
There are numerous infrared hot-spots and massive young stellar objects 
(Urquhart et al. 2008), ultra compact (UC) \HII\ regions and molecular masers 
located on the periphery of the \HII\ region. The integrated radio flux is 
indicative of $>30$ deeply embedded 07V stars (Clark \& Porter 2004).
The detection of these high-mass star formation tracers, within the boundary 
layer between the ionized and molecular gas, would suggest that, not only are 
high-mass stars currently forming, but their formation may have been triggered 
through the interaction of the ionization front and surrounding molecular 
material.

A multi-wavelength observing program was designed to study the star 
formation across the G305 complex. High frequency radio continuum observations 
are a key ingredient of this program as they allow the large scale structure 
of the ionized gas to be traced, and the emission from very young high-mass 
stars to be identified via their compact and ultra-compact \HII\ regions. 
Previous cm continuum observations have either covered small areas at high 
resolution, e.g., around the methanol masers listed in Walsh et al. (1998), or 
have been of low angular resolution and sensitivity (Danks et al. 1984, Caswell 
\& Haynes 1987, Haverkorn et al. 2006). Mapping large Galactic star forming 
regions on the scale of G305 at high frequency and sub-arcminute resolution 
has only recently been made possible with the commissioning of CABB on the 
ATCA. This is primarily due to the factor of at least 16 increase in frequency 
bandwidth, which effectively increases mapping speeds by the same factor 
over the previous system to reach the same sensitivity. A further speed 
increase is gained by the potential to map two frequencies simultaneously 
using the two IFs.

G305 was mapped using ATCA with CABB at two frequencies (centred around 5 and 
9~GHz), requiring 357 separate mosaic points to cover the whole region. 
Multiple array configurations were required to obtain sufficient $uv$-coverage 
to recover both extended and compact structures. In total, six different 
configurations were used to obtain fairly uniform baseline coverage between 
30~m and 6~km, which provides sensitivity to angular scales of a few arcseconds 
to a few tens of arcminutes. The observations were made in snapshot mode, i.e.
$6-8 \times 10$ second integrations were spent on each of the 357 pointings. 
This was repeated for each configuration; the total integration time for the 
project was 75-h.

In Fig.~\ref{G305_midIR} we present a three colour mid-infrared image of G305 
overlaid with contours of the integrated 6-cm emission obtained from the three
ATCA hybrid arrays (H75, H168 and H214); these configurations primarily consist
of short baselines and so are very sensitive to the large scale structure of 
the ionized gas. The ionized gas is correlated with prominent emission 
structures seen in the mid-infrared, tracing the interface between the \HII\ 
region and surrounding molecular gas. Combining the radio emission with 
tracers of molecular gas (e.g., Hindson et al. 2010) and dust will allow 
investigating the impact of massive stars associated with Danks~1 and 2 on 
the structure of G305's molecular gas as well as its influence on current 
and future star formation in the region. 

\begin{figure} 
 \mbox{\psfig{file=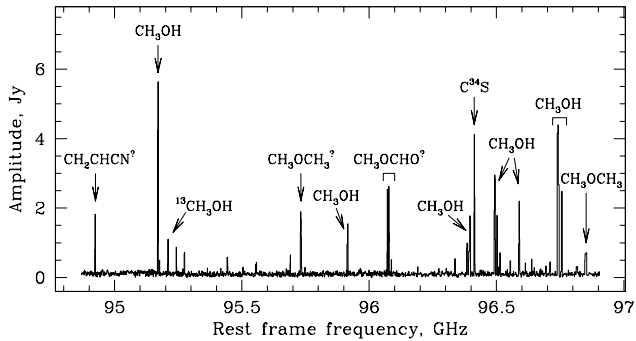,width=8.5cm}}
\caption{\label{ngc6334spectrum}
   CABB 3-mm broadband molecular line spectrum of the hot-core chemistry 
   forest associated with NGC~6334\,I (see \S~6.2). The spectrum extends from 
   about 95 to 97~GHz; more than 30 molecular lines are easily detected.}
\end{figure}

\subsection{A forest of molecular lines from NGC~6334\,I} 

NGC~6334, also known as the `Cats Paw Nebula', is a giant molecular cloud /
\HII\ region located at a distance of $\sim$1.7~kpc in the southern Galactic 
Plane. Situated in its northern part is the high-mass star-forming site known 
as NGC~6334\,I. This site contains a hot molecular dust core at the head of a 
bright cometary UC\,\HII\ region and is associated with a wealth of young 
high-mass star-forming phenomena such as masers, molecular outflows, a forest 
of molecular line hot-core chemistry as well as several compact millimetre 
continuum sources (e.g. Beuther et al. 2007, Hunter et al. 2006). Walsh et al. 
(2010) studied a $5\arcmin \times 5\arcmin$ region encompassing NGC~6334\,I and 
another bright (sub)millimetre continuum source further north (NGC~6334\,I(N)) 
at $\lambda$ 3~mm (from 83.5 to 115.5~GHz) using the 22-m Mopra telescope 
(angular resolution $\sim$36\arcsec). They detect a total of 52 transitions 
from 19 species (molecules, ions and radicals). Beuther et al. (2008) studied 
NGC~6334\,I with the ATCA in 2006 (pre-CABB) at around 88.4~GHz, achieving an 
rms of 23 mJy\,beam$^{-1}$. While the angular resolution was 
high ($2\farcs3 \times 1\farcs6$), the modest observing bandwidth of 128~MHz 
only permitted two molecular emission lines to be detected simultaneously. 

NCG~6334\,I was again observed on 16 April 2009, as part of the CABB science 
commissioning program. Antenna array configuration H168 was used with an 
on-source observing time of 1.6-h. Fig.~\ref{ngc6334spectrum} shows the 2~GHz 
wide-band CABB spectrum (95 -- 97~GHz) towards the central position of 
NGC~6334\,I at $\alpha,\delta$(J2000) = 
$17^{\rm h}\,20^{\rm m}$\,53\fs35, --35\degr\,47\arcmin\,1\farcs6. More than 
30 molecular lines are easily detected simultaneously, demonstrating the power 
of the Compact Array Broadband Backend (CABB) for multi-transition chemical 
line studies. 

\subsection{Galactic methanol maser observations}  

\begin{figure} 
 \mbox{\psfig{file=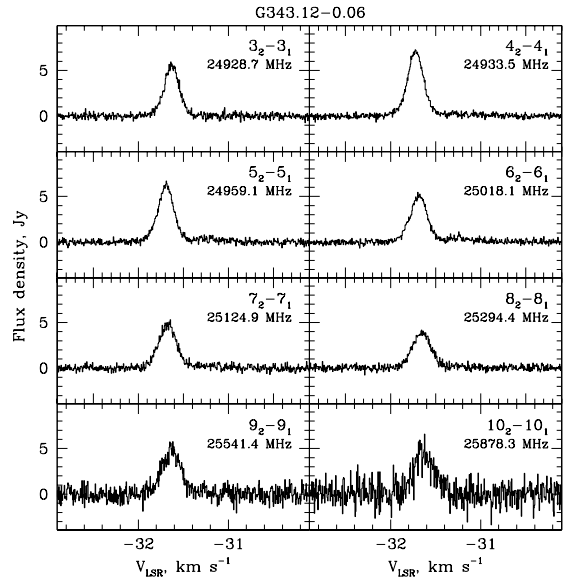,width=8cm,angle=0}}
\caption{\label{g343zoom}
   Methanol maser lines in the high-mass star-forming region G343.12--0.06 
   around 25~GHz, as observed with eight CABB 1~MHz {\em zoom} windows. The 
   full width of each {\em zoom} window is 12\kms\ and the velocity resolution 
   is 6\ms.}
\end{figure}

Fig.~\ref{g343zoom} demonstrates the power of CABB with its first zoom modes
(obtained on 3 Jun 2010, by Maxim Voronkov). Eight zoom windows were chosen 
to cover bright methanol maser transitions near 25~GHz (J$_2-$J$_1$~E methanol 
series) in the high-mass star-forming region G343.12--0.06 (IRAS~16547--4247).
This source has a jet-driven molecular outflow embedded in a molecular cloud
(Brooks et al. 2003). The interaction between the outflow and the cloud 
produced a number of masers across the region (Voronkov et al. 2006). The CABB
observations shown here replicate in a more efficient way most of the 25~GHz 
observations from Voronkov et al. (2006) which were obtained with the original
ATCA correlator by cycling through all the transitions sequentially (a stronger
J=10 transition was observed instead of a weak J=2 maser reported in the 
original study). Each zoom window was 1~MHz wide (about 12\kms\ of velocity
coverage) and had 2048 spectral channels, providing a spectral resolution of
about 6\ms. Even the narrow maser lines were notably oversampled, as the 
spectral resolution provided by this CABB configuration was a factor of
4 to 8 times better than that of  Voronkov et al. (2006). 

\begin{figure} 
\begin{tabular}{c}
 \mbox{\psfig{file=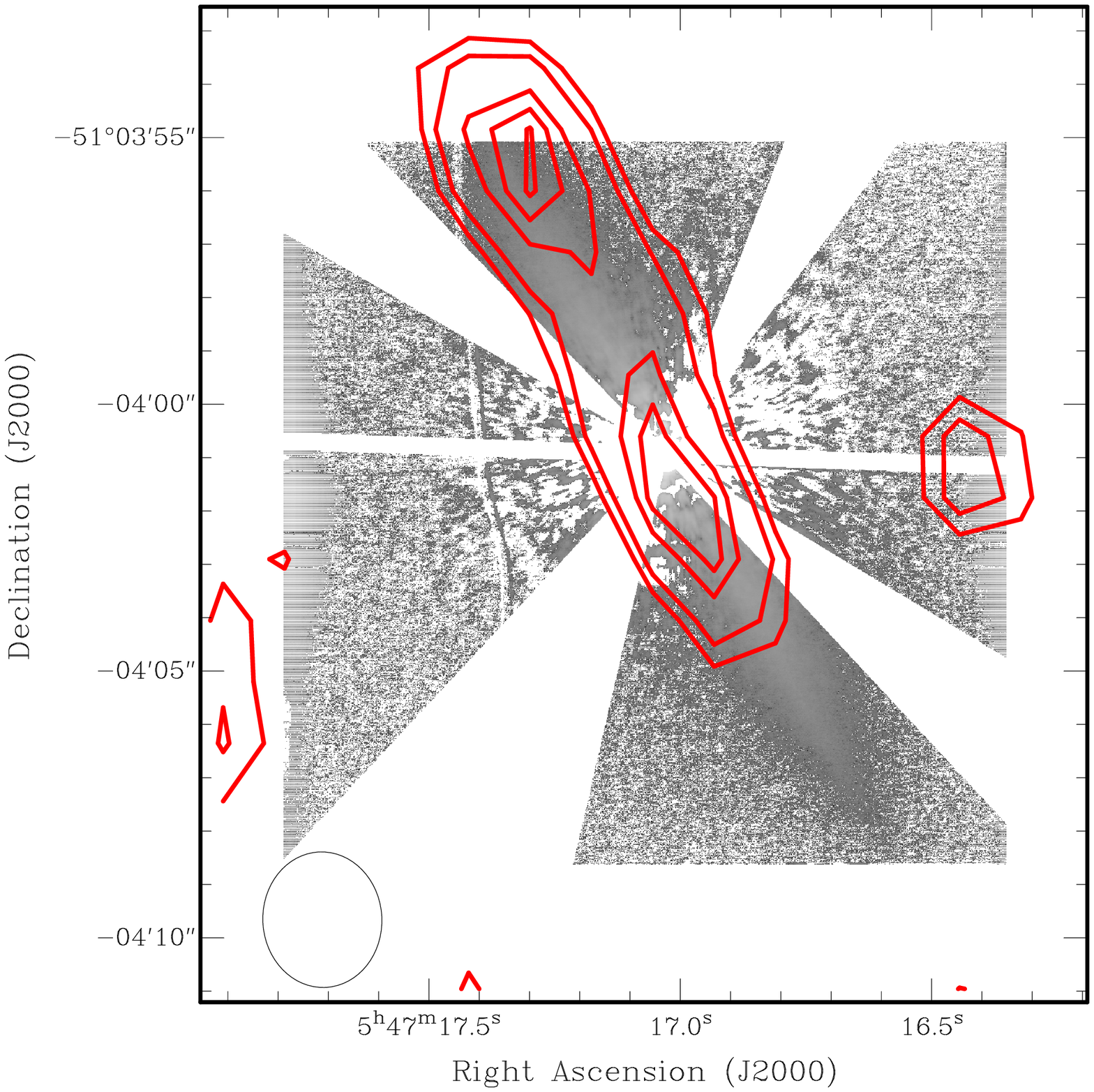,width=7cm,angle=0}} \\
 \mbox{\psfig{file=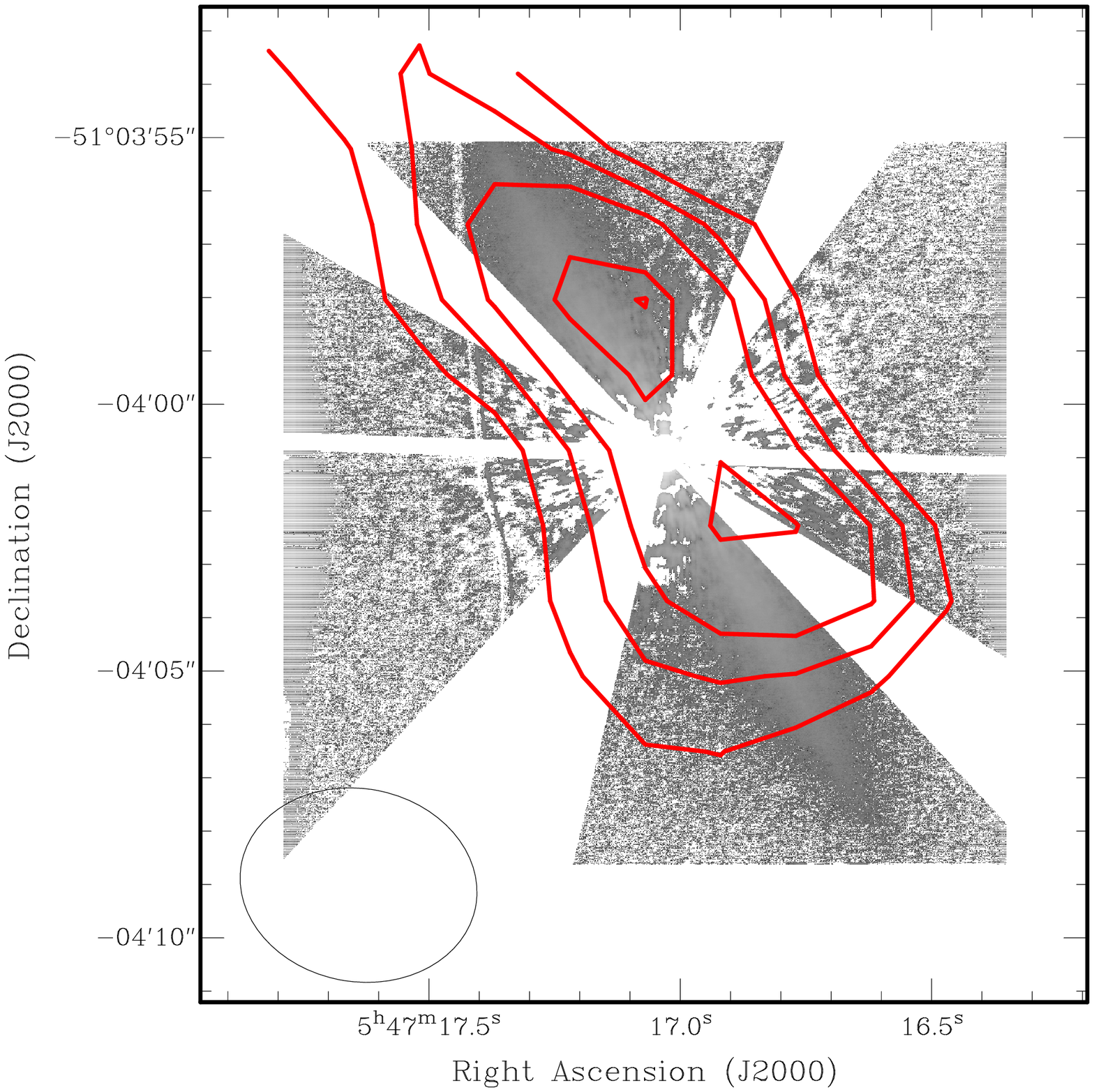,width=7cm,angle=0}} 
\end{tabular}
\caption{\label{betapic}
   The $\beta$\,Pictoris debris disk (at a distance of 19.5~pc 1\arcsec\
   corresponds to 20~AU). Contours show the extended radio continuum emission 
   from the dust grains in the disk at 3-mm {\bf (top)} and 7-mm {\bf (bottom)}
   overlaid onto an ESO/NACO $H$-band (1.65$\mu$m) image in greyscale 
   (Boccaletti et al. 2009). The radio data were taken with the ATCA in the 
   H214 configuration, using CABB with 4~GHz of bandwidth. The contour levels 
   are 1.5, 2, 3, 3.5 and 4$\sigma$ (where $\sigma$ = 95~$\mu$Jy is the r.m.s. 
   noise) at 3-mm (beam $\approx$ 2\farcs5) and 2, 3, 4, 5, and 6$\sigma$ 
   (where $\sigma$ = 21.8~$\mu$Jy) at 7-mm (beam $\approx$ 4\farcs5). The NE 
   peak appears brighter than the SW peak in both images (see \S~6.4).}
\end{figure}

\subsection{The $\beta$ Pictoris Debris Disk} 
 
Debris disks represent the final evolutionary phase of protoplanetary disks 
around young stars. They are optically thin, contain little or no gas and are 
generally thought to be in the final planet forming stage (Wyatt 2008). At a 
distance of just 19.3~pc (Crifo et al. 1997), the $\beta$~Pictoris debris 
disk is an excellent target for high resolution imaging. 

$\beta$~Pictoris has been extensively studied since its discovery as an 
infrared excess star by IRAS (Aumann 1984) and subsequent imaging of its
dusty disk by Smith \& Terrile (1984). The latter revealed a nearly edge-on 
disk $\sim$400~AU (20\arcsec) in extent. At an age of 10--20~Myrs 
(Zuckerman et al. 2001), $\beta$~Pic is mature enough to host planets and, 
indeed, the direct detection of an $\sim$8 Jupiter mass planet was recently 
reported (e.g., Lagrange et al. 2010).

Modelling of the thermal and scattered light by Artymowicz (1997) indicates 
that the $\beta$~Pic disk contains a wide range of grain sizes from sub-micron 
through to at least millimetre dimensions. Simple order of magnitude estimates 
show that even the smaller dust grains cannot be primordial because they would 
be blown away radiatively on timescales of tens of orbits. Collisions between 
larger bodies will lead to fragmentation, since there is very little gas to 
dampen the relative grain velocities (Lecavelier des Etangs et al. 2001), and 
thus these collisions act to replenish the dust. 
Debris disks can have a very different appearance at different wavelengths. 
Sub-mm and mm observations trace the largest grains, which are least affected 
by stellar radiation, suggesting that these grains are the best probe for the 
larger parent bodies responsible for the dust production. 

Millimetre interferometry provides high angular resolution and much needed 
information on the largest grain populations in debris disks. The close 
proximity of $\beta$~Pic and its southern declination make it an ideal target 
for mm-imaging with the ATCA. 

Using CABB 3-mm (92 -- 96~GHz) and 7-mm (42 -- 46~GHz) observations with the 
ATCA in the H214 configuration, Maddison \& Wright (2011) were able to clearly
resolve the debris disk of $\beta$\,Pic (see Fig.~\ref{betapic}). These are 
the longest wavelength detections of any debris disk obtained to date. To 
achieve the 4~GHz wide bandwidth, they placed the two 2~GHz IF bands adjacent 
to each other; integration times were 8-h at 3-mm (12/13 Sep 2009) and 6.5-h 
at 7-mm (13/14 Sep 2009). Maddison \& Wright (2011) found clear evidence of 
structure within the dust and a direct detection of inner disk clearing, in 
agreement with recent Sub-Millimetre Array (SMA) 1.3-mm observations by Wilner 
et al. (2011). 

\begin{figure*} 
 \mbox{\psfig{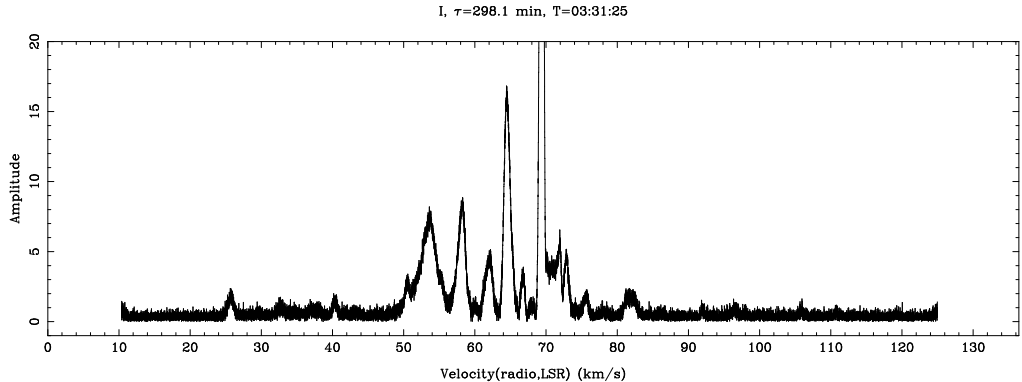}} 
\caption{\label{sagb2zoom16}
  High velocity resolution CABB spectrum of water masers in Sgr\,B2, taken on 
  17 Dec 2010 (uncalibrated). Here $16 \times 1$~MHz {\em zoom} channels were 
  concatenated, i.e., overlapped in 0.5~MHz steps, to give 17,408 channels 
  over 8.5~MHz bandwidth; the velocity resolution is 7\ms. Note that at the 
  outermost edges of the concatenated band \tsys\ gradually rises by less 
  than a factor two.}
\end{figure*}
  
\subsection{Water masers in the Sgr\,B2} 
 
In Dec 2010 the first CABB observations with $16 \times 1$~MHz {\em zoom} 
channels were successfully demonstrated. Fig.~\ref{sagb2zoom16} shows the 
high-velocity resolution line data of water masers in Sgr\,B2. Here the {\em 
zoom} channels are overlapped in 0.5~MHz steps to obtain a uniform spectrum 
over 8.5~MHz of bandwidth with a velocity resolution of 7\ms. McGrath et al. 
(2004) show a similar spectrum at 0.66\kms\ velocity resolution, while Reid 
et al. (2009) determined the trigonometric parallax of Sgr\,B2 by measuring 
the H$_2$O maser positions with the Very Large Baseline Array (VLBA) over one 
year.
	
\subsection{The Circinus Galaxy} 

\begin{figure} 
 \mbox{\psfig{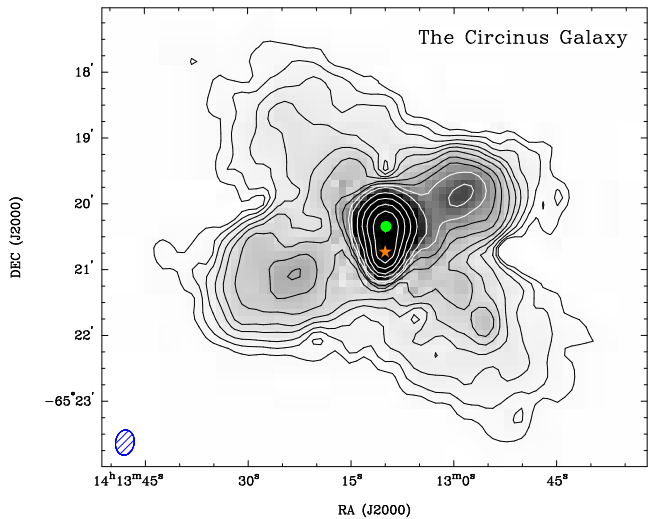}} 
\caption{\label{circinus}
   ATCA 6-cm radio continuum emission of the nearby Circinus Galaxy (observed 
   on 2 Apr 2009, EW352 array, CABB 2~GHz bandwidth). We detect the active 
   galactic nucleus (AGN; marked with a green dot), the recent supernova 
   (SN1996cr; marked with an orange star), the inner galaxy disk ($PA$ = 
   223\degr) and the orthogonal radio lobes ($PA$ = 115\degr). Contour levels 
   are 0.3, 0.5, 0.8, 1.3, 1.8, 2.6, 3.8, 5, 7, 10, 20, 40, 80, 160, and 320 
   mJy\,beam$^{-1}$ (white contours start at 7 mJy\,beam$^{-1}$). The 
   synthesized beam ($23\arcsec \times 17\arcsec$) is displayed in the bottom 
   left corner. For more information see \S~6.6.}
\end{figure}

The Circinus Galaxy is, at a distance of 4~Mpc, one of the closest active 
galaxies. Discovered relatively recently (Freeman et al. 1977) and obscured 
by Galactic foreground stars and dust, it nevertheless remains a prominent 
observing target at a large range of frequencies. It harbours an active 
galactic nucleus (AGN) surrounded by a star-forming ring, an extended stellar 
disk, double radio lobes (orthogonal to the disk), and an enormous \HI\ gas 
envelope (Jones et al. 1999; Curran, Koribalski \& Bains 2008). Bauer et al. 
(2008) discovered a supernova, named SN1996cr, located just 25\arcsec\ south 
of the AGN, adding to the continuing interest in this spectacular galaxy (see
Fig.~\ref{circinus}). 

CABB 6-cm observations were carried out in April 2009 using the EW352 array. 
The 2~GHz band, centred at 5500~MHz, was divided into 2048 channels. During the 
9-h synthesis we regularly observed the nearby calibrator PKS\,1352--63. The 
bandpass and amplitude calibrator PKS\,1934--638 was observed at the start of 
the session. Radio interference was detected in numerous channels, particularly
around 5600~MHz (bandwidth 6~MHz) probably related to the 5625~MHz weather 
radar near the town of Moree. Our goal was to reproduce (or enhance) the 6-cm 
radio continuum map of the Circinus galaxy obtained by Elmouttie et al. (1998),
to study the emission from the nuclear region, the disk and the radio lobes. 
Their published ATCA data consisted of $5 \times 12$-h with a bandwidth of 
128~MHz, reaching an r.m.s. of 120~$\mu$Jy\,beam$^{-1}$. With an effective 
increase of a factor 20 in the useable bandwidth ($20 \times 100$~MHz), a 
factor $\sim$2 in \tsys, highly improved uv-coverage and dynamic range, we 
estimate a theoretical noise of $\sim$13~$\mu$Jy\,beam$^{-1}$ for the CABB 
6-cm imaging of the Circinus galaxy ('uniform' weighting).
The data were calibrated in {\sc miriad} by splitting the 2~GHz band into $64 
\times 32$~MHz channels, then imaged and further calibrated in {\sc difmap}
(Shepherd 1997) using a combination of model-fitting and self-calibration. 
The resulting image (uniformly weighted; no self-calibration) is shown in 
Fig.~\ref{circinus}. In the nuclear region the AGN and the supernova SN1996cr 
are clearly detected (and well-resolved in the high-resolution image, not 
displayed here). Thanks to CABB, the extended, star-forming disk of the 
Circinus galaxy and its spectacular radio lobes are revealed in detail (the
image dynamic range is $\sim$1000:1). Our CABB 6-cm image closely resembles 
the pre-CABB 20-cm image by Elmouttie et al. (1998). An in-depth analysis of 
the gas distribution and star formation in Circinus is under way (For, 
Koribalski \& Jarrett 2011).

\begin{figure} 
 \mbox{\psfig{file=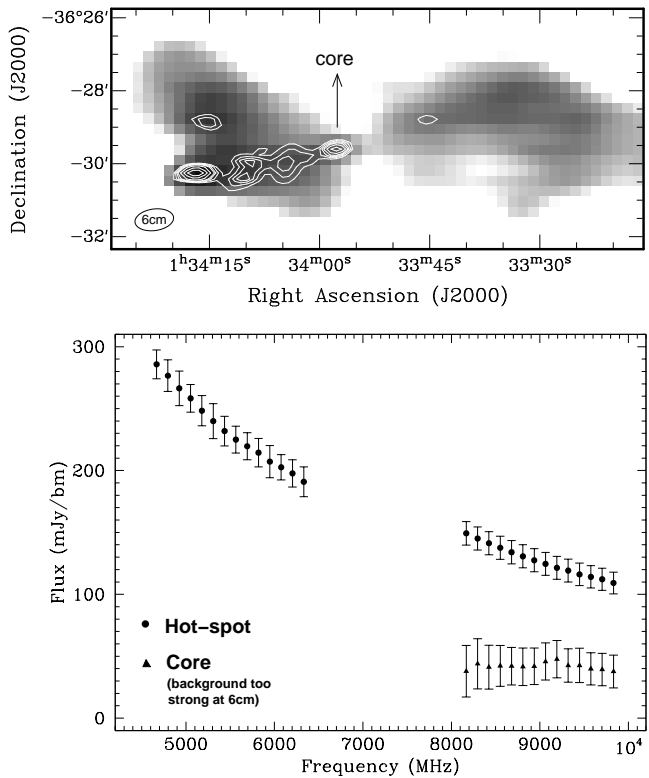,width=8.5cm,angle=0}} 
\caption{\label{ngc612}
   The radio galaxy NGC~612 (PKS\,0131--36) as observed at 3- and 6-cm during 
   CABB commissioning in April 2009 with the ATCA in the compact H168 antenna
   configuration. {\bf Top:} preliminary radio continuum image at 6-cm 
   (greyscale) and 3-cm (contours). The contour levels are 4, 7, 10, 13, 20, 
   30, and 40 mJy\,beam$^{-1}$. {\bf Bottom:} spectral index of the eastern 
   hot-spot (dots) as well as the nucleus (triangles) across the full $2 
   \times 2$~GHz range of the two bands. The frequency separation between 
   consecutive data points reflects the 128~MHz bandwidth of the original 
   ATCA correlator. The spectral index properties are visualised using linear 
   rather than the commonly used logarithmic axes (see \S~6.7 for details).}
\end{figure}

\subsection{The Radio Galaxy NGC~612} 

NGC~612 (PKS\,0131--36) is one of the nearest powerful radio galaxies ($z$ =
0.0297; e.g., Ekers et al. 1978). Its eastern radio jet shows a hot-spot, 
typical for Fanaroff \& Riley type-II radio sources.
The host galaxy is a dust-lane S0 with a young stellar population detected
along its disk (V\'eron-Cetty \& V\'eron 2001; Goss et al. 1980; Holt et al.
2007). NGC~612 also contains a large-scale disk of \HI\ gas (140~kpc in 
diameter) as well as a faint bridge-like \HI\ structure ($\sim$400~kpc in 
size) that stretches towards a gas-rich companion (Emonts et al. 2008). 
The proximity of NGC~612 allows us to have a close look into the physical
processes associated with powerful radio galaxies.

The complex and bright structure of the radio continuum source (roughly the 
size of the primary beam at 6-cm) made NGC~612 an excellent target to image 
during CABB commissioning. On 9 and 10 April 2009, NGC~612 was the first 
science target observed simultaneously in the 3- \& 6-cm bands, using the 
H168 hybrid configuration. Over the following months, these observations 
were repeated in H75 and H168 at three additional epochs. The main aims were 
to perform wide-band imaging across the primary beam, derive instantaneous 
spectral index information of both the nucleus and hot-spot across a large 
frequency range and investigate CABB's performance for deriving polarization 
properties of this source.

Figure~\ref{ngc612} shows preliminary results obtained from the CABB 
observations of NGC~612. It includes total intensity images of its radio 
continuum emission at 3- \& 6-cm and a spectral index diagram (over 4~GHz) of 
both its nucleus and eastern hot-spot. For the latter, each individual data 
point was obtained by extracting 128~MHz sub-bands from the full CABB 2~GHz 
band. Standard reduction was performed on each 128~MHz sub-band (i.e. 
bandpass/phase/flux calibration, imaging, cleaning and primary beam correction 
-- see \S~5 for details). We then measured the integrated flux densities of 
the nucleus and the eastern hot-spot by fitting Gaussians functions in each 
`128~MHz' sub-image.
The steep spectral index of the eastern radio hot-spot\footnote{In 
   Fig.~\ref{ngc612} the spectral index of NGC~612's eastern hot-spot is 
   plotted using linear rather than logarithmic axes in order to better 
   visualise the instantaneous frequency coverage of CABB.}
is in stark contrast to the rather flat spectrum of the nucleus (at 6-cm the 
continuum of the inner radio lobes becomes too bright to clearly distinguish 
the resolved nucleus). These results are in agreement with continuum 
observations at 20~GHz by Burke-Spolaor et al. (2009). The wealth of spectral 
index information that can be obtained with a single CABB observation, once 
the calibration is finalised, is striking. The preliminary analysis also 
indicates that the eastern hot-spot is highly polarised ($\sim$25\%) at 6-cm.

\subsection{Deep continuum observations of the ECDFS} 

The Extended Chandra Deep Field South (ECDFS) is part of the Australia 
Telescope Large Area Survey (ATLAS; Norris et al. 2006). The goal of ATLAS is 
to study the origin and evolution of both star-forming and AGN-dominated
galaxies by making deep multi-wavelength radio observations complemented by
deep infrared and optical imaging as well as spectroscopy.

\begin{figure} 
 \mbox{\psfig{file=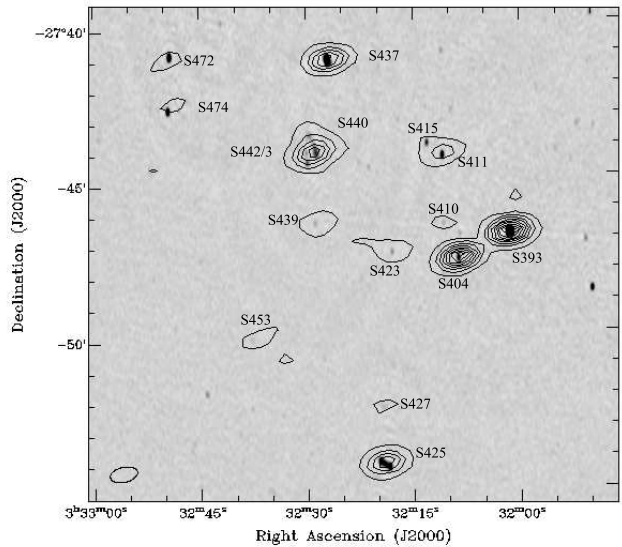,width=9cm}} 
\caption{\label{ecdfs}
   Preliminary CABB radio continuum image of the Extended Chandra Deep Field 
   South (ECDFS) at 6-cm (contours) overlaid onto a pre-CABB 20-cm image 
   (greyscale) from Hales et al. (2011). Contours are at intervals of 
   $\approx$100\,$\mu$Jy. The 20-cm data were taken with the ATCA in a 6-km 
   configuration, while the 6-cm data were observed in the compact H168 
   array.}
\end{figure}

The ECDFS was observed for about 4-h in the ATCA H168 configuration on 19 
Apr 2009. The 2~GHz CABB band was centred at 5500~MHz and divided into 2048 
channels. The data were processed in {\sc miriad}, initially using the full 
2~GHz spectrum. Without self-calibration, an r.m.s. noise of 23~$\mu$Jy was 
obtained (see Fig.~\ref{ecdfs}); this was improved to 16~$\mu$Jy (close to 
the theoretical r.m.s.) after one iteration of phase self-calibration. 
Re-processing and primary beam correction in smaller sub-bands (e.g. 128~MHz) 
will be required to obtain correct flux densities and spectral indices for 
all sources in the field.

Source S415, located several arc minutes from the pointing centre, is an 
example of the rare "Infrared-faint Radio Sources" published by Norris et al.
(2006); see also Beswick et al. (2008). Using ATLAS data, Middelberg et al. 
(2011) give its 20-, 13- and 6-cm flux densities as $2.64 \pm 0.54$~mJy, 
$0.70 \pm 0.08$~mJy, and $<$0.33~mJy, respectively, resulting in a spectral 
index of $\alpha_{20,13} = -2.38 \pm 0.43$. The CABB 6-cm data reveal a 
$\approx$80~$\mu$Jy radio source at the position of S415, confirming the steep 
spectral index. This is important since such sources are often associated with 
high-redshift radio galaxies (Krolik \& Chen 1991; de Zotti et al. 2010). 
S415 has no detected 3.6~$\mu$m counterpart to a limit of $\sim$1~$\mu$Jy, 
giving it a $S_{\rm 20cm}/S_{\rm FIR}$ ratio of $\sim$2000, compared to 
$\sim$200 typical for radio-loud quasars and $\sim$10 for star-forming 
galaxies. 

\subsection{High redshift CO} 

A crucial component in the study of star formation throughout the Universe is 
a detailed knowledge of the content and properties of the molecular gas. An 
excellent tracer for molecular gas is the emission-line of carbon monoxide 
(CO), which is strong enough to be observed out to high redshifts. However, 
since the first detection of CO at $z>2$ by Brown $\&$ Vanden Bout (1991), 
observations of molecular gas at high redshifts have been limited by the fact 
that, at the high observing frequency of CO ($\nu_{\rm rest} = 115$~GHz for 
the lowest rotational transition), existing narrow-band receivers covered a 
velocity range that was generally not much larger than the CO line width or 
the accuracy of the redshift. Moreover, the lack of millimetre receivers in
the 20--50~GHz regime resulted in a focus on observations of the higher order 
rotational transitions. While the higher order transitions are good tracers 
for the molecular gas that is dense and thermally excited, large reservoirs 
of low-density and widely distributed gas could be hiding in the lower 
rotational CO transitions (e.g. Papadopoulos et al. 2000, 2001).

In its coarsest spectral line mode, the $2 \times 2$~GHz CABB bands with 1~MHz 
spectral resolution are ideal for targeting high-$z$ CO lines, in particular 
regarding the lower order transitions that are shifted into the ATCA 7-mm and 
15-mm bands. For example, a source at $z$ = 3.8 can be observed in CO(1--0) 
at 24~GHz (15-mm band) as well as CO(2--1) at 48~GHz (7-mm band), with 
observations of the latter covering a velocity range of more than 12,000\kms\
per 2~GHz band at a resolution of $\sim$6\kms.

The first detection of high-$z$ CO with CABB was made by Coppin et al. (2010), 
who detected the CO(2--1) transition in a distant sub-millimetre galaxy at 
$z$ = 4.76, indicating the presence of $\sim$1.6 $\times 10^{10}$\Msun\ of 
molecular gas. Recently, Emonts et al. (2011a) detected the CO(1--0) transition
in a powerful radio galaxy at $z \sim 2$, representing $\sim$5 $\times 
10^{10}$\Msun\ of molecular gas. This is only the third known high-$z$ radio 
galaxy detected in CO(1--0), after detections in two others at a significantly 
lower signal-to-noise levels (Greve et al. 2004; Klamer et al. 2005). These 
initial results show the enormous advantage CABB delivers for ATCA studies of 
molecular gas and star formation in the Early Universe. The excellent 
performance of CABB for high-$z$ CO observations is demonstrated by Emonts et 
al. (2011b), including a detailed description of their observing strategy
for millimetre spectral line work with CABB. 

\subsection{Polarization synthesis}  

The large frequency bandwidths that have become available with CABB are also 
useful for radio polarization observations, which can tell us about the 
properties of magnetic fields in the Milky Way and beyond. In particular 
Faraday rotation, an effect where the plane of polarization of a radio wave 
is rotated when the wave passes through an ionised region with an embedded 
magnetic field, has been a very popular method over many years (Gaensler et 
al. 2005; Brown et al. 2007). Faraday rotation strongly varies with the 
observing wavelength ($\propto \lambda^2$), which means that the polarization 
vectors at different frequencies will have different orientations. The 
magnitude of this variation with frequency (known as the `rotation measure', 
or RM) depends on the magnetic field strength and the free electron density 
in the column that the radio wave passes through. From the measured RM we 
can estimate the magnetic field strength, if we can correct for the electron 
density contribution to RM.

A large observing bandwidth gives a great improvement in sensitivity, also 
in polarization, but only if we can correct for the change in orientation 
of the polarization vectors with frequency, which is the result of the 
Faraday effect. By simply summing the different frequency channels some of 
the polarised signal will be lost if the polarization vectors have different 
orientations. The novel technique of rotation measure synthesis (or `RM 
synthesis'; Brentjens \& De Bruyn 2005\footnote{RM synthesis was developed 
   independently in 1996 by De Bruyn (NFRA note 655) and by Killeen et al. 
   (1999).}) helps us out: if the Faraday 
effect rotates the polarization vectors as a function of frequency, then by 
choosing the correct RM for that line of sight the polarization vectors 
can be de-rotated (in RM synthesis-speak the RMs are referred to as 
`Faraday depths'). This way the polarization vectors from all the frequency 
channels in the observing band can be combined, and the noise level is thus 
greatly reduced. Typically one would use many trial Faraday depths and 
calculate how much polarised flux is recovered at each Faraday depth, to find 
the Faraday depth of the polarised emission for a particular line of sight. 
Furthermore, when there are regions along the line of sight that emit 
polarised radiation with different Faraday depths (e.g. when there is 
additional Faraday rotation between the emission regions), then each 
Faraday depth will produce a polarization vector that rotates at a specific 
rate as a function of observing frequency. RM synthesis can separate such
regions, since it is essentially a Fourier transform. This way RM synthesis 
adds a depth dimension to measurements of the Faraday effect.

\begin{figure*} 
 \mbox{\psfig{file=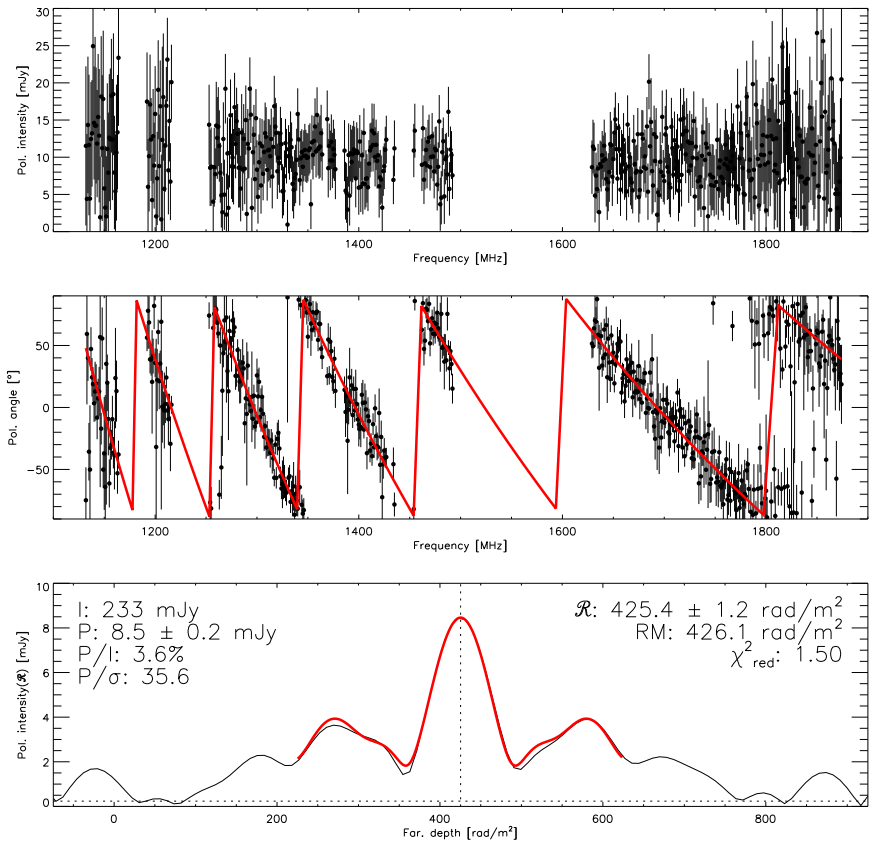,width=12cm}} 
\caption{\label{cabb-pol}
   The polarization properties of an extragalactic source on a line of sight 
   close to the Galactic supershell GSH287+04--17 (source coordinates: 
   $\alpha,\delta$(J2000) = $10^{\rm h}\,48^{\rm m}\,07\fs653$, 
      --53$^\circ$\,13\arcmin\,21\farcs38). 
   The top and middle panel show the polarised intensity and polarization angle 
   as a function of frequency, which were derived from the observations. The 
   red line in the middle panel indicates the best-fitting RM = 426 rad/m$^2$ 
   (the fit has a reduced $\chi^2$ of 1.5). The bottom panel shows how much 
   power (polarised intensity) is contributed by each Faraday depth. In this 
   case 8.5~mJy is generated at a Faraday depth of 425 rad/m$^2$, which matches
   the average polarised intensity from the top panel and the best-fitting RM 
   from the middle panel well. The solid red line indicates the profile that 
   can be expected from RM synthesis on the basis of the frequency coverage of 
   the data. The bottom panel illustrates the strength of RM synthesis: 
   although polarised intensity is detected in each channel at about the 
   2$\sigma$ level (top panel), combining all frequency channels, which is 
   possible with RM synthesis, gives a detection with a combined polarised 
   signal/noise level of 36$\sigma$ (bottom panel; the noise level in the 
   bottom panel is about 0.2 mJy).}
\end{figure*}

To study whether magnetic fields are important for the evolution of the 
Galactic supershell GSH287+04--17 we used CABB 20-cm data in combination 
with RM synthesis. For this purpose we determined Faraday 
depths of extragalactic sources that are seen through the wall of the 
supershell, and we compared these to the Faraday depths of extragalactic 
sources that lie just outside the shell wall. Fig.~\ref{cabb-pol} shows
the polarised intensities and polarization angles that were derived for one 
such source and the polarised intensity that is recovered for different trial 
Faraday depths. In this example the extragalactic source is the only source 
of polarised emission along the line of sight, and the expected response 
from RM synthesis (similar to the dirty beam in synthesis imaging), which 
is indicated with a thick red line, closely matches the measured response. 
We determined the Faraday depths of $\sim$100 extragalactic sources that 
are seen through or close to the ionised inner part of the supershell wall.
The enhanced free electron density will naturally lead to an increase in RM 
if the magnetic field remains the same. However, we did not find such an 
enhancement of the RMs of the extragalactic sources here, but we did find a 
narrow filament that runs nearly parallel to the Galactic plane, where the 
RM varies by $\ga$250 rad/m$^2$. This filament spans the entire width of the 
area we surveyed (and is therefore $\ga$5\degr\ wide) but it is only 1\degr\ 
high. Since there appears to be no enhancement in the H$\alpha$ intensity 
that is associated with this ribbon it must be magnetic in origin, and it 
will be very interesting to investigate how such a magnetized structure can 
have survived in the Galactic interstellar medium.

\subsection{AT20G quasar study} 

Mahony et al. (2011) observed a sample of over 1100 QSOs to study their radio 
luminosity distribution, a subject that has long been debated in the literature
(e.g., Kellermann et al. 1989, Cirasuolo et al. 2003). QSOs are often separated
into two categories; `radio-loud' and `radio-quiet', but the dividing line 
between these two classes remains unclear. The increased bandwidth of CABB, 
and hence increased sensitivity compared to the original ATCA correlator, 
allows a large sample of objects to be observed in a smaller amount of time, 
whilst also probing further into the `radio-quiet' regime. A more detailed 
description of the observing program is given in (Mahony et al. 2010b).
Targets were selected from the RASS--6dFGS catalogue (Mahony et al. 2010a); a 
catalogue of 3405 AGN selected from the ROSAT ALL Sky Survey (RASS) Bright 
Source Catalogue (Voges et al. 1999) that were observed as part of the 6dF 
Galaxy Survey (6dFGS; Jones et al. 2009). The observations were carried out
in snapshot mode using a hybrid array (H168) on the ATCA from 2008 to 2010, 
using a two-step process. Firstly, all sources were observed for $2 \times 40$
seconds reaching an r.m.s. noise of $\sim$0.2~mJy. Secondly, non-detections 
were re-observed for $2 \times 5$ minutes, reaching an r.m.s. noise of 
$\sim$0.1~mJy. 

\begin{figure} 
 \mbox{\psfig{file=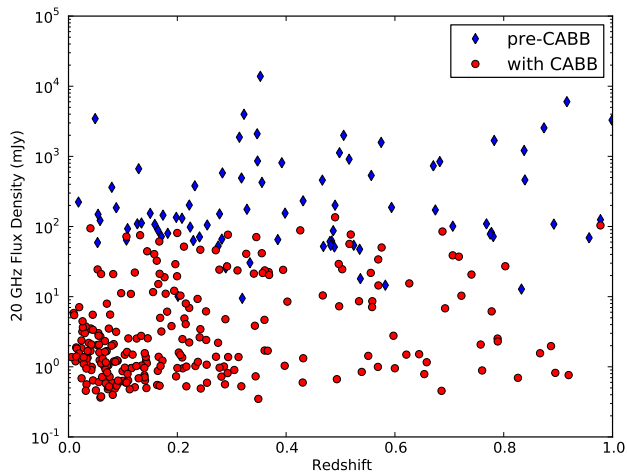,height=7cm}} 
\caption{\label{mahony}
   ATCA 20~GHz flux densities for a sample of X-ray selected QSOs against 
   redshift. Blue diamonds indicate targets that were observed using the 
   original correlator on the Compact Array, while red circles show the 
   sources observed more recently using CABB. The data were obtained between 
   2008 and 2010 in the H168 configuration.}
\end{figure}

Preliminary results are shown in Fig.~\ref{mahony} which compares the 20~GHz 
flux densities (vs redshift) of the X-ray selected sample achieved pre-CABB 
and with CABB (since March 2009). It highlights the significant increase in 
sensitivity that CABB has achieved in comparison to the original correlator.

\subsection{Surprising S-Z results for Two Clusters} 

Pre-CABB, no Sunyaev-Zeldovich Observations (S-Z) observations with resolution 
better than a few arc minutes were possible.  Even for rich clusters of 
galaxies many days of integration were required to marginally detect the low 
brightness S-Z dip.

This all changed with the increased sensitivity provided by the CABB system.  
Two independent observations of different galaxy clusters made with CABB at 
18~GHz detected the S-Z dip with sub arc minute resolution and in both 
observations the S-Z was displaced from the peak of the X-ray emission from 
the cluster gas. 
  
\begin{figure} 
 \mbox{\psfig{file=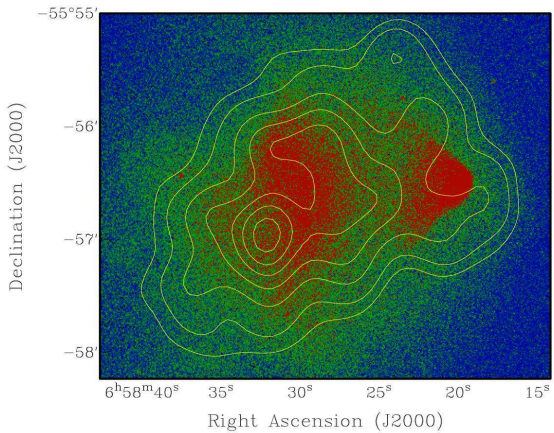,width=8cm}} 
\caption{\label{bullet}
   ATCA 18~GHz image of the S-Z effect in the Bullet cluster (Malu et al. 
   2011) in contours overlaid onto a Chandra X-ray emission map displayed in
   color. The hottest gas (deepest SZE feature) is displaced from the densest 
   regions traced by the X-ray emission (see \S~6.12). The ATCA observations 
   were made with CABB in April 2009 (H168 array; $2 \times 12$-h) and June 
   2009 (H75 array; $3 \times 8$-h) using both IFs, i.e., $2 \times 2$~GHz 
   bands. --- The Chandra X-ray Observatory is operated by the Smithsonian 
   Astrophysical Observatory on behalf of NASA.}
\end{figure}

Massardi et al. (2010) observed CL J0152--1357, a massive cluster at $z$ = 
0.83, and concluded that the cluster was still recovering from its merging 
formation; it is not in virial equilibrium and must contain lower temperature 
high pressure gas not detected in X-ray observations.
Malu et al. (2011) observed the Bullet cluster (1E0657--56) at $z$ = 0.296.  
They note that the deepest S-Z features seem to avoid the regions of most 
intense X-ray emission (see Fig.~\ref{bullet}). These two independent results 
support Malu et al.'s assertion that modeling cluster dynamics is non trivial 
and that our lack of understanding of cluster merger astrophysics may be 
limiting our ability to model the cosmological distribution of S-Z counts. 

\subsection{Pulsar binning} 

CABB is capable of operating in pulsar mode, where a pulse profile can be 
recorded every cycle for each frequency channel for each polarization in up to 
two IFs. Typically 32 phase bins across the pulsar period can be recorded with 
a minimum bin time of $\sim$110$\mu$s. 

\begin{figure} 
  \mbox{\psfig{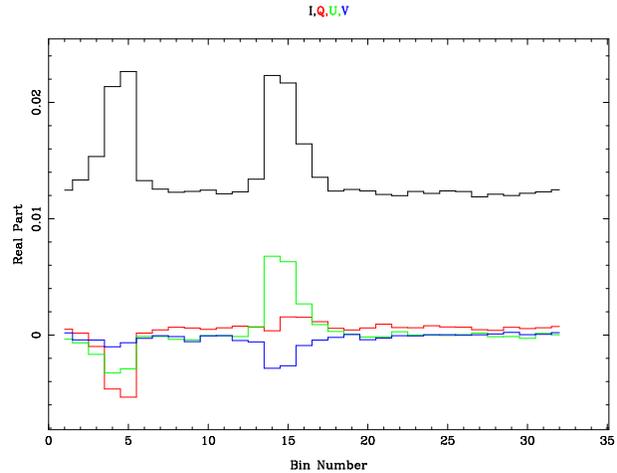}}
\caption{\label{pulsar-binning}
  Observations of the pulsar PSR\,1259--63 made on 12 Jan 2011 at a central 
  frequency of 9.0~GHz and a total bandwidth of 2~GHz. The pulse profile is 
  shown in 32 phase bins across the pulse period of $\sim$48~ms. The total 
  intensity trace (black) shows the characteristic double peaked profile for 
  this pulsar (Manchester \& Johnston 1995). The red and green traces show 
  Stokes Q and U, respectively; the pulsar is highly linearly polarised. 
  Circular polarization is shown in blue.}
\end{figure}

Observations in pulsar mode require knowledge of the pulsar rotation's period 
at the observatory and how the period changes over the course of the 
observation. This is done using a {\sc tempo2} predictor file (Hobbs et al. 
2009) and requires that the pulsar ephemeris details are stored in the pulsar 
catalogue.  If this is the case then the code "psrpred" can be executed on 
the correlator computer to set up the predictor file. 

Various science applications for pulsar mode can be conceptualised. One example 
is to determine where pulsars emit during the so-called off-pulse phase or 
whether the emission there is truely zero. A further application would be the 
removal of the pulsar signal to determine whether there is underlying emission 
from a pulsar wind nebula (Stappers, Gaensler \& Johnston 1999). Finally, 
transient continuum (unpulsed) emission from a pulsar system can also be 
detected and the flux density measured without contamination from the pulsar 
signal (Johnston et al. 2005).

The periastron passage of the pulsar PSR\,B1259--63 around its companion Be 
star is known to produce transient emission over a wide range of wavelengths, 
from the radio to the TeV. The most recent periastron occurred on 14 Dec 2010,
and a major campaign was mounted to maximise the coverage over the 
electromagnetic spectrum. In the radio, observations were made at the ATCA 
on a regular basis from early Nov 2010 until Mar 2011 using CABB in 
pulsar mode. Fig.~\ref{pulsar-binning} shows an observation made at 9~GHz 
on 11 Jan 2011. 
The two pulses which make up this pulsar's profile are clearly seen sitting 
on a pedestal of transient emission with a flux density of $\sim$12~mJy. The 
transient emission is unpolarized. However, the pulsar itself is highly 
polarized (including circular polarization) as shown by the lower traces. 
Full analysis of this data is still in progress.

\subsection{SEDs for AT20G sources observed by Planck} 
\begin{figure} 
 \mbox{\psfig{file=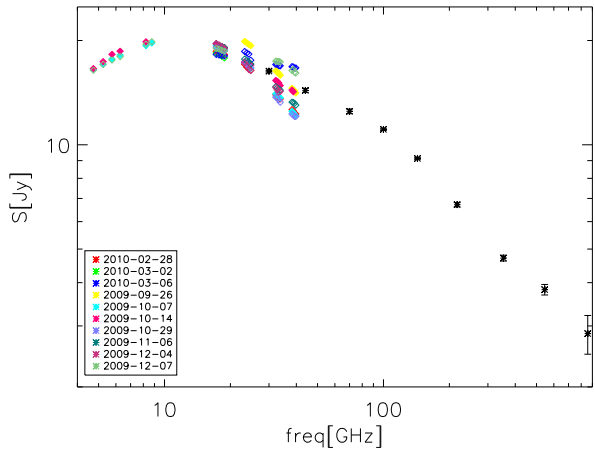,height=9cm,angle=90}} 
\caption{\label{massardi}
  Spectral energy distribution of the qusasar PKS\,B1921--293 as measured by 
  the ATCA (4.5 -- 40~GHz) over several epochs (marked with coloured open 
  diamonds) as part of the PACO project (Massardi et al. 2011b) and by Planck 
  released in the ERCSC (30 -- 545~GHz; black asterisks, Planck collaboration 
  2011). Note that the displayed Planck flux densities were obtained from maps 
  integrated over all scans during the satellite first sky survey (i.e. Aug 
  2009 -- Mar 2010). Filled diamonds indicate the ATCA observing epochs within 
  10 days from a Planck scan at frequencies $<$100 GHz.}
\end{figure}

The Australia Telescope 20~GHz (AT20G) Survey (Murphy et al. 2010; Massardi et 
al. 2011a) used a custom-made analogue correlator, with 8~GHz of bandwidth in 
dual polarization, on three antennas and was completed while the CABB system 
was under construction. Although it had similar bandwidth to CABB, it had no 
delay tracking, so could not be used to follow-up sources after the survey was 
completed. CABB follow-up observations of 483 AT20G sources provided flux 
density measurements almost simultaneously with Planck observations (Massardi
et al. 2011b). These extend the frequency coverage of Planck providing for 
the first time the instantaneous Spectral Energy Distributions (SED) of the 
brightest AGN from 4.5 to 40~GHz (ATCA) and from 30 to 857~GHz (Planck). The 
project is known as the "Planck-ATCA Co-eval Observations" (PACO). As an 
example we show the SED of PKS\,B1921--293 (also known as OV--236 or here 
AT20G J192451--291430), a redshift 0.35 QSO and a very bright flat spectrum 
AGN (see Fig.~\ref{massardi}), often used as ATCA bandpass calibrator at 
mm-wavelengths. The wide-bandwidth multi-frequency CABB data samples a large 
fraction of the SED, up to 40~GHz, and shows continuity with the Planck Early 
Release Compact Source Catalogue (ERCSC, Planck Collaboration 2011) data at 
higher frequencies. The multi-epoch CABB data highlight the importance of near 
simultaneous observations to sample the full SED without the distortion due 
to variability.

\section{Summary}

The Compact Array Broadband Backend (CABB) upgrade has been described and 
its potential for new discoveries demonstrated. The high flexibility of CABB
allows a large range of oberving modes, providing wide bandwidth as well as
high spectral resolution and full polarization output. The increased bandwith, 
effectively by a factor 20 (from 
$\sim$100~MHz of useable bandwidth in the original correlator to the full
CABB 2~GHz bandwidth), available in two IF bands (dual polarization) or four 
IF bands (single polarization) together with multi-bit digitisation and
much improved instantaneous $uv$-coverage (for continuum observations) has 
substantially advanced the science capabilities of the ATCA. Furthermore,
the widening and combination of the original, relatively narrow 20-cm and 
13-cm bands to one broad band spanning 1.1 to 3.1~GHz allows multi-line (e.g.,
searching for \HI\ and OH emission and absorption lines) plus continuum 
observations over a large range of velocities. We hope that you, the reader,
will be tempted to observe with the new system if you have not already done 
so.

\section*{Acknowledgements}
The CABB project is grateful for the funding received through the Australian
Government's Major National Research Facilities (MNRF) 2001 program. Our
appreciation also goes to the engineers and technicians at CSIRO's Marsfield 
and Narrabri sites whose dedicated efforts made for the successful design, 
construction and implementation of this system. We thank the editor and the
referee for a delightful report, improving the content and clarity of the 
paper.

\end{document}